\documentclass[12pt]{article}
\usepackage{a4wide}
\usepackage{amssymb}
\usepackage[tbtags]{amsmath}
\usepackage{exscale}

\begin{document}
{\renewcommand{\thefootnote}{\fnsymbol{footnote}}
\hfill  IGC--11/12--2\\
\medskip
\begin{center}
{\LARGE  Deformed General Relativity and Effective Actions from Loop Quantum Gravity}\\
\vspace{1.5em}
Martin Bojowald\footnote{e-mail address: {\tt bojowald@gravity.psu.edu}}
 and George M.\ Paily\footnote{e-mail address: {\tt gmpaily@phys.psu.edu}}
\\
\vspace{0.5em}
Institute for Gravitation and the Cosmos,
The Pennsylvania State
University,\\
104 Davey Lab, University Park, PA 16802, USA\\
\vspace{1.5em}
\end{center}
}

\setcounter{footnote}{0}

\newcommand{\lP}{\ell_{\mathrm P}}
\newcommand{\md}{{\mathrm{d}}}
\newcommand{\tr}{\mathop{\mathrm{tr}}}
\newcommand{\sgn}{\mathop{\mathrm{sgn}}}
\newcommand{\stmet}{{\bf g}}

\newcommand{\be}{\begin{equation}}
\newcommand{\ee}{\end{equation}}
\newcommand{\dif}{\mathrm{d}}
\newcommand{\fund}[2]{\ensuremath{\frac{\delta #1}{\delta #2}}}
\newcommand{\pard}[2]{\ensuremath{\frac{\partial #1}{\partial #2}}}
\newcommand{\met}[2][{ij}]{\ensuremath{g_{#1}(#2)}}
\newcommand{\metd}[1]{\ensuremath{g_{#1}}}
\newcommand{\ph}[1]{\Phi^{#1}}
\newcommand{\bph}[1]{\bar{\Phi}^{#1}}
\newcommand{\phd}[2]{\ensuremath{\Phi^{#1}_{#2}}}
\newcommand{\phdl}[1]{\Phi^{#1}_{|l}}
\newcommand{\bphdl}[1]{\bar{\Phi}^{#1}_{|l}}
\newcommand{\um}[1]{g^{#1}}
\newcommand{\mom}[2][{ij}]{\ensuremath{\pi^{#1}(#2)}}
\newcommand{\ex}[2][{ij}]{\ensuremath{v_{#1}(#2)}}
\newcommand{\vv}{\ensuremath{(x \leftrightarrow x')}}
\newcommand{\momd}[3][{ij}]{\ensuremath{\pi^{#1}_{#2}(#3)}}
\newcommand{\Gb}{\ensuremath{L^{ij}(x)}}
\newcommand{\Gbd}[1]{\ensuremath{L^{ij}_{#1}(x)}}
\newcommand{\ao}{\ensuremath{\beta^{\emptyset}}}
\newcommand{\aod}[1]{\ensuremath{\beta^{\emptyset}_{#1}}}
\newcommand{\aan}[1]{\ensuremath{\beta^{{#1}}}}
\newcommand{\aad}[2]{\ensuremath{\beta^{{#1}}_{#2}}}
\newcommand{\kd}[2]{\ensuremath{\delta^{#1}_{#2}}}
\newcommand{\abn}[1]{\ensuremath{\beta^{{#1}}}}

\newcommand{\kp}{\ensuremath{K_\varphi}}
\newcommand{\kx}{\ensuremath{K_x}}
\newcommand{\ep}{\ensuremath{E^\varphi}}
\newcommand{\erp}{\ensuremath{(E^{x})'}}
\newcommand{\epp}{\ensuremath{(E^{\varphi})'}}
\newcommand{\gp}{\ensuremath{\Gamma_\varphi}}
\newcommand{\dyx}{\ensuremath{\delta(y,x)}}
\newcommand{\dxy}{\ensuremath{\delta(x,y)}}
\newcommand{\dxz}{\ensuremath{\delta(x,z)}}
\newcommand{\dyxp}{\ensuremath{\delta'(y,x)}}
\newcommand{\dxyp}{\ensuremath{\delta'(x,y)}}
\newcommand{\dyxpp}{\ensuremath{\delta''(y,x)}}
\newcommand{\jig}[1]{\,\ensuremath{{}^{#1}\!H}}

\begin{abstract}
Canonical methods can be used to construct effective actions from
deformed covariance algebras, as implied by quantum-geometry
corrections of loop quantum gravity. To this end, classical
constructions are extended systematically to effective constraints of
canonical quantum gravity and applied to model systems as well as
general metrics, with the following conclusions: (i) Dispersion
relations of matter and gravitational waves are deformed in related
ways, ensuring a consistent realization of causality. (ii) Inverse-triad
corrections modify the classical action in a way clearly
distinguishable from curvature effects. In particular, these
corrections can be significantly larger than often expected for
standard quantum-gravity phenomena. (iii) Finally, holonomy corrections
in high-curvature regimes do not signal the evolution from collapse
to expansion in a ``bounce,'' but rather the emergence of the universe
from Euclidean space at high density. This new version of signature-change
cosmology suggests a natural way of posing initial conditions, and a
solution to the entropy problem.
\end{abstract}

\section{Introduction}

A major consequence expected for quantum gravity is the emergence of
non-classical space-time structures such as discrete or
non-commutative ones. Any such modification by quantum properties
deforms the standard notion of covariance and thus gives rise to
possible new actions and interaction terms. Developments in this
direction are of interest for a fundamental understanding of space and
time, and also for potential observations of quantum gravity:
Unexpected structures may give rise to new effects, or magnify
others. One example is early-universe cosmology. Assuming the
classical space-time structure with the usual notion of covariance
results in higher-curvature terms in an effective action, and only
small quantum corrections are possible, suppressed by factors of
$\lP/\ell_{\cal H}$ of the Planck length by the Hubble
distance. Non-classical space-time structures, on the other hand, can
sometimes circumvent such limitations and magnify expected effects
compared to what standard higher-curvature terms would deliver (as
realized explicitly in \cite{InflConsist,InflTest}).

However, relaxing conditions on covariance in a consistent way is not a
straightforward task. Space-time properties such as discreteness or
non-commutatitivy are often obtained at some kind of kinematical quantum level
far removed from direct space-time analysis. One may, for instance, look at
operators that quantize geometrical quantities such as distances or areas, or
analyze the behavior of test particles or, mathematically, test functions on
quantum space-time. These concepts are not directly related to the dynamics of
space-time itself, and so it is initially not clear what form of deformed
covariance principle could be used to formulate dynamics on such modified
space-times and to find the possible correction terms analogous to
higher-curvature effective actions. 

Fortunately, an abstract but powerful substitute exists in canonical
formulations: Any generally covariant theory in four space-time dimensions has
a gauge algebra of four local generators per space-time point, which serve as
constraints on suitable initial values and generate space-time transformations
on phase-space functions by canonical transformations. If quantization leads
one to modified expressions for these generators, covariance is realized ---
albeit perhaps deformed --- if the generators still obey an algebra of the
classical dimension. From the perspective of general gauge theory, the same
number of spurious degrees of freedom is then removed by the constraints as
classically, and all equations of motion derived for the system are guaranteed
to be consistent. The theory is anomaly-free.

More specifically, in generally covariant theories there are three smeared
constraints per point labeled by spatial vector fields, the diffeomorphism
constraint $D[N^i]$ depending on an arbitrary shift vector $N^i$, and a fourth
one labeled by a function, the Hamiltonian constraint $H[N]$ depending on the
lapse $N$. Classically, these phase-space functions obey the
hypersurface-deformation algebra\footnote{In addition to $D[N^i]$ and $H[N]$,
  there are primary constraints given by the momenta of the non-dynamical $N$
  and $N^i$. Their algebra just mimics the canonical structure, not space-time
  structure, and can thus be ignored for the purposes of this article.}
\begin{eqnarray} \label{Algebra}
  \{D[N^i],D[M^j]\}&=& D[{\cal L}_{M^j}N^i]\,\\
  \{H[N],D[N^j]\} &=& H[{\cal L}_{N^j}N]\, \label{HD}\\
  \{H[N_1],H[N_2]\} &=& D[g^{ij}(N_1\partial_jN_2-
  N_2\partial_jN_1)] \label{HHclass}
\end{eqnarray}
with the spatial metric $\met{x}$.  (In this article we denote the metric on a
spatial 3-manifold in space-time by $\met{x}$, and by \mom{x} its conjugate
momentum, using for the sake of easier comparison the notation of the articles
\cite{LagrangianRegained,Regained} which we will follow closely in some
parts. For an overview of canonical methods, the reader is referred to
\cite{CUP}.)

Gauge transformations $\delta F= \{F,H[N]+D[N^i]\}$ of a phase-space function
$F$ then agree with the changes implied by infinitesimal deformations of the
spatial hypersurfaces in space-time. In a passive picture, this gauge
transformation agrees with a coordinate change along a space-time vector field
$\xi^{\mu}$ with components given in terms of the spatial fields $N$ and $N^i$
(see e.g.\ \cite{LapseGauge}). The whole hypersurface-deformation algebra
presents a large extension of the local Poincar\'e algebra, which is recovered
for linear $N$ and $N^i$ in a local coordinate patch \cite{KucharHypI}. A
general property of the algebra is that it is largely insensitive to the
dynamics of the underlying covariant theory: all higher-curvature theories
have constraints obeying the same algebra; see e.g.\ \cite{HigherCurvHam} for
an explicit calculation. This uniqueness statement can be reversed if the
derivative order of one's theory is constrained to be at most two in the
equations of motion, in which case the form of the action (up to the values of
Newton's and the cosmological constant) can uniquely be recovered from the
hypersurface-deformation algebra \cite{LagrangianRegained,Regained}. Mimicking
the usual tensorial arguments to fix the terms of the Einstein--Hilbert
action, the dynamics, to the lowest order of derivatives, is thus uniquely
determined by the algebra of constraints.

The algebra itself is rather rigid as well, making it difficult to
implement new covariance principles and correction terms other than
higher-derivative ones. A new result of recent years, however, is that
loop quantum gravity, if it can be consistent at all, gives rise to
modified hypersurface-deformation algebras. With different kinds of
quantum corrections characteristic of the theory, this has been seen
for perturbative inhomogeneity
\cite{ConstraintAlgebra,VectorHol,ScalarHol}, in $2+1$ dimensions
\cite{ThreeDeform,TV} and in spherically symmetric models
\cite{SphSymmPSM,JR,LTBII}. Different physical consequences for
cosmology \cite{ScalarGaugeInv,LoopMuk,InflConsist} and for properties
of black holes \cite{ModCollapse,ModifiedHorizon,ModCollapse2} have
resulted. As a common form of the modified constraint algebra, one can
write
\begin{equation} \label{HH}
 \{H_{(\beta)}[N_1],H_{(\beta)}[N_2]\}= D[\beta
 g^{ij}(N_1\partial_jN_2-N_2\partial_jN_1)]
\end{equation}
in terms of a phase-space function $\beta[g_{ij},\pi^{ij}]$ determined by the
quantum corrections considered. Poisson brackets in (\ref{Algebra}) involving
the diffeomorphism constraint remain unmodified (except in the case of
\cite{VectorHol} which has been superseded by \cite{ScalarHol}). 

That a closed algebra still arises is far from trivial, and shows that general
relativity, at least in the models considered, can be deformed
consistently. The systems obtained correspond to a more general form of
space-time covariance than usually taken into account.\footnote{Such
  deformations are similar in spirit to doubly special relativity
  \cite{DSR1,DSR2,Rainbow,DSR}, but the two different concepts are not
  straightforwardly related: Doubly special relativity deforms the Poincar\'e
  algebra non-linearly, with corrections depending on algebra generators such
  as the energy. In (\ref{HH}), the Poincar\'e algebra is affected as well by
  the subalgebra of the hypersurface-deformation algebra with linear $N$ and
  $N^i$ in Cartesian coordinates, but correction functions $\beta$ depend on
  phase-space variables $g_{ij}$ and $\pi^{ij}$, not on the constraints as
  algebra generators. In some backgrounds, a relationship can nevertheless be
  established, as will be discussed elsewhere.}  In this article, we will
assume an algebra of the form (\ref{HH}) and analyze what the possible
consequences for action principles are. With action principles at hand, the
interpretation of deformed constraint algebras will become more
intuitive. Moreover, they provide manifestly covariant (in the deformed sense)
formulations of the underlying models of loop quantum gravity from which the
quantum corrections have been extracted.

The conclusions we will be able to derive are surprisingly rich: (i)
We will obtain a clear separation of some corrections from others. In
particular, inverse-triad corrections in loop quantum gravity will
play a much more characteristic role than holonomy corrections of the
same theory, or higher-curvature corrections of general form. (ii) The
dynamics of loop quantum gravity near a spacelike classical
singularity takes on a specific form in which spatial derivatives
become subdominant. A scenario similar to but more generic than the
BKL picture follows. (iii) Loop quantum gravity will be seen to give
rise to signature change in strong-curvature regimes. This new feature
of the theory, overlooked so far in minisuperspace models, gives rise
to new and improved cosmological scenarios.

\section{Overview of deformed constraint algebras in loop quantum gravity}
\label{s:Overview}

Canonically, the quantum effects of interacting gravitational
theories, often expressed by higher-curvature effective actions,
are derived from quantum back-reaction \cite{Karpacz}: While expectation
values of semiclassical states follow nearly the classical
trajectories, additional state parameters such as fluctuations and
other moments influence the quantum trajectory. Coupled equations of
motion for expectation values and the moments can, in some regimes of
adiabatic nature, be reformulated as the usual equations of low-energy
effective actions \cite{EffAc}. 

Obviously, these effects should play a large role for quantum gravity and
cosmology.  But in addition to the ubiquitous quantum back-reaction (or
corrections from loop diagrams in perturbative terms), there are
characteristic quantum corrections expected for loop quantum gravity,
providing two distinct quantum-geometry effects: (i) higher powers of spatial
curvature components (intrinsic and extrinsic) stemming from the appearance of
holonomies of the Ashtekar--Barbero connection instead of direct connection
components in quantized constraints \cite{RS:Ham,QSDI}, and (ii) natural
cut-off functions of divergences of factors containing inverse components of
the densitized triad, arising from spatial discreteness \cite{QSDI,QSDV}. The
first type of quantum-geometry corrections is usually referred to as
``holonomy corrections,'' the second as ``inverse-triad corrections'' (or, in
the context of nearly isotropic cosmology, ``inverse-volume
corrections''). Both can be expanded as series of corrections by components of
{\em spatial} tensors in the constraints, not by scalar invariants of
space-time tensors as one is used to from covariant effective actions. Neither
the reconstruction of an action principle from the constraints nor properties
of covariance are obvious in such a situation, and the only systematic way to
determine such features is an analysis of the constraint algebra. As shown in
several model systems so far, the hypersurface-deformation algebra is
generically deformed by quantum-geometry. In particular, corrections cannot be
written purely as higher-curvature terms added to the Einstein--Hilbert
action, as often expected for quantum gravity. One of the main questions to be
addressed in this article is what actions and covariance properties could be
realized instead.\footnote{Sometimes in models of loop quantum gravity,
  higher-curvature actions have been used as an ansatz to compare with
  quantum-geometry corrections in restricted contexts
  \cite{ActionRhoSquared,ActionLovelock}. However, the gauge-fixings or
  complete reductions to homogeneity employed to formulate consistent
  equations in such a procedure leave too many ambiguities and prevent
  sufficient access to the gauge content of the theory. A large class of
  corrections in homogeneous or gauge-fixed models is possible which would be
  ruled out by a consistent extension to inhomogeneity; and for a given
  corrected version of a homogeneous model, many different action principles
  can be found by such an analysis. They would all yield the same homogeneous
  equations, but differ uncontrollably regarding the dynamics of
  inhomogeneities.}

In this section we summarize the models investigated so far for their
properties of deformations of the constraint algebra, split into the
two types of quantum-geometry corrections. (Quantum back-reaction has
not yet been analyzed to completion in this context, but the procedure
would follow \cite{EffAc,EffCons,EffConsRel}.) The set of models in
which consistent deformations have been achieved is quite diverse, but
the general form of $\beta$ appears to be insensitive to model
specifications. The constraint algebra therefore displays universal
implications for covariant space-time structure.

\subsection{Inverse-triad corrections}

In loop quantum gravity, space-time geometry is described by canonical fields
$A_i^I$ and $E^i_I$, a connection related to curvature and the densitized
triad, instead of the spatial metric $g_{ij}$ and its momentum
$\pi^{ij}$. These fields have advantages for a background independent
quantization because they can be smeared without reference to an auxiliary
metric structure: The connection is integrated along curves $e$ in space to
obtain holonomies $h_e(A)={\cal P}\exp(\int_e \tau_IA^I_i\dot{e}^i{\rm
  d}\lambda)$, and the densitized triad, dual to a 2-form, is integrated to
fluxes $F_S(E)=\int_S\tau^IE^i_In_i{\rm d}^2y$ through surfaces $S$ in
space. Here, $\tau_I=\frac{1}{2}i\sigma_I$ are generators of su(2), related to
the Pauli matrices. The canonical structure $\{A_i^I(x),E^j_J(y)\}= 8\pi
\gamma G\delta_i^j \delta^I_J \delta(x,y)$ with the Barbero--Immirzi parameter
$\gamma$ \cite{Immirzi,AshVarReell} provides a regular relation for
$\{h_e(A),F_S(E)\}$ free of delta functions.

Holonomies and fluxes are promoted to basic operators of the resulting
quantum theory, and they represent the canonical fields in all
composite operators such as Hamiltonians. Both types of basic
operators imply some form of non-locality because they are integrated
rather than pointlike fields. Using holonomies for connection
components, moreover, implies that there are higher-order corrections
when the exponential is expanded, compared with classical expression
which are usually polynomials of degree at most two in the
connection. Fluxes also give rise to corrections in addition to their
non-locality: They are quantized to operators with discrete spectra,
containing zero as an eigenvalue. Such operators are not invertible,
and yet an inverse of the densitized triad (or its determinant) is
needed to quantize matter Hamiltonians (usually in the kinetic part) and
the Hamiltonian constraint. Well-defined operators with inverse
densitized triad components as their classical limit do exist
\cite{QSDI}, but they have strong quantum corrections for small values
of the fluxes. Correction functions, obtained from expectation values
of inverse-triad operators \cite{QuantCorrPert}, then primarily depend
on the fluxes, or on the densitized triad and the spatial metric. In
non-Abelian situations, there can also be some dependence on the
connection via higher-order terms \cite{DegFull}.

Inverse-triad corrections cannot easily be formulated consistently in
homogeneous models, where the rescaling freedom of the scale factor under
changes of coordinates may be broken unless one properly refers to underlying
discreteness scales. However, with some inhomogeneous input consistent
formulations exist \cite{Consistent,InflObs,InflTest} and show the importance
of these quantum-geometry corrections.  Quantization of the dynamics can
proceed only if a substitute for the non-existing inverse of an elementary
flux $\hat{F}$ is found, which according to \cite{QSDI} is possible by using
Poisson-bracket identities. If we write schematically\footnote{We
  use U(1)-expressions $\exp(i\delta A)$ instead of SU(2)-holonomies in this
  equation, as is often realized in symmetric models. With non-Abelian
  SU(2)-holonomies in the full theory, the holonomies in Poisson brackets or
  commutators may not completely cancel depending on the ordering, leaving
  also a dependence on $A$. But the leading term in an $A$-expansion will
  remain unchanged which, as we will see later, is the crucial contribution
  from inverse-triad corrections.}  $|F|^{q-1}{\rm sgn}(F) = (8\pi G\gamma
\delta q)^{-1}i\exp(i\delta A)\{\exp(-i\delta A),|F|^q\}$ with a connection
component $A$, we have an inverse of $F$ on the left-hand side for $q<1$ while
the right-hand side can be quantized without requiring an inverse of $F$ if
$q>0$. The Poisson bracket can straightforwardly be quantized: There is an
operator $\hat{F}$ whose positive power $|\hat{F}|^q$ can easily be taken via
the spectral decomposition. While loop quantum gravity does not provide an
operator for $A$, it does have well-defined quantizations of ``holonomies''
$h=\exp(i\delta A)$.  Finally, we turn the Poisson bracket into a commutator
divided by $i\hbar$, and achieve to quantize $|F|^{q-1}{\rm sgn}(F)$ in spite
of the non-existence of an inverse of $\hat{F}$. 

The resulting operator is well-defined and has an inverse power of $F$
as its classical limit, approached on the part of the spectrum of
$\hat{F}$ with large eigenvalues.  There are quantization ambiguities
which prevent one from finding a unique correction function
\cite{Ambig,ICGC}. The typical form, however, follows from algebraic
properties and results in $\widehat{|F|^{q-1}{\rm sgn} F}=
\left(|\hat{F}+\Delta F|^q- |\hat{F}-\Delta F|^q\right)/2q\Delta
F$ with a Planckian $\Delta F\approx \ell_{\rm P}^2=\hbar G$. Such
corrections with a tiny Planck area may seem small, but
$\langle\hat{F}\rangle$ as a fundamental flux or area of a discrete
state is typically Planckian as well.  For small flux values,
characteristic quantum corrections result \cite{InvScale},
constituting inverse-triad corrections. We collect inverse-triad
effects in a correction function
\begin{equation}
 \bar{\alpha}(F)= |F|^{1-q}{\rm sgn}F\cdot \langle
     \widehat{|F|^{q-1}{\rm sgn}F}\rangle_{F}= |F|^{1-q} {\rm sgn}F
     \frac{|F+\Delta F|^q- |F-\Delta F|^q}{2q\Delta F}+ \mbox{moment
     terms}
\end{equation}
up to $\hat{F}$-fluctuations and higher moments, using an expectation value in
a state peaked at flux $F$.

\subsubsection{Cosmology}

The most general class of models shown so far to have a consistently deformed
constraint algebra is the one of perturbative inhomogeneity around spatially
flat Friedmann--Robertson--Walker models \cite{ConstraintAlgebra}, including
inverse-triad corrections. In this case, $\beta=\bar{\alpha}^2$ in (\ref{HH})
with the background function $\bar{\alpha}$ of inverse-triad corrections
depends on the scale factor $a$. These corrections, as in the full theory,
arise because loop quantum cosmology \cite{LivRev,Springer,SIGMA} quantizes
the scale factor, or rather its square $|p|=a^2$ equipped with a sign for
spatial orientation, to an operator $\hat{p}$ with discrete, equally spaced
spectrum. The spectrum contains zero as an eigenvalue, and therefore $\hat{p}$
has no densely defined inverse.

Apart from their formal derivation, inverse-triad corrections in
cosmology are characterized by cut-off effects of classically
diverging quantities such as $a^{-1}$. For degenerate geometries, or
near the big-bang singularity of isotropic models, discreteness
effects lead to non-divergent quantities when shift-operators
$\exp(i\delta A)$ instead of differential operators $-i\hbar
\partial/\partial F$ are used in the commutators of inverse-triad
operators. For fluxes in isotropic space-times, we write
$F=\ell_0^2a^2$ with the coordinate size $\ell_0$ of elementary
plaquettes in a regular-lattice discrete state (choosing $F$ to be
positive without loss of generality). The cut-off behavior is clearly
visible from properties of the ratio
\begin{equation}
 \frac{\bar{\alpha}(\ell_0^2a^2)}{a^{2(1-q)}}= \frac{|\ell_0^2a^2+\Delta
   F|^q-|\ell_0^2a^2-\Delta F|^q}{2q\Delta F}
\end{equation}
which approaches zero (instead of infinity) for $a\to 0$, and asymptotes to
the classical expression $1/a^{2(1-q)}$ for $a\gg a_*$ well above a
characteristic scale $a_*=\sqrt{\Delta F}/\ell_0$. The latter depends on the
discreteness behavior of an underlying state, which is responsible for the
quantum correction and the implicit cut-off of divergences associated with
$1/a^{2(1-q)}$. Regarding the scaling behavior of $a$ and $a_*$, the
background behavior of inverse-triad corrections, as derived in
\cite{InvScale,QuantCorrPert}, has been made consistent in inhomogeneous
settings in \cite{Consistent,InflTest}. (The precise form of $\bar{\alpha}$ as
a function of phase-space variables will not be important in this article.)

For perturbative inhomogeneities in spatially flat isotropic models, a
consistent deformation (\ref{HH}) results at least if $\bar{\alpha}$
is close to one \cite{ConstraintAlgebra}. Once it is ensured that the
algebra of constraints closes, several consistency conditions for the
correction functions arise. The background behavior of $\bar{\alpha}$
appearing in the gravitational part of the Hamiltonian constraint
remains unrestricted, but analogous correction functions in possible
matter contributions must be related to it and are no longer
completely arbitrary. The case of a scalar field will be discussed in
more detail below, Section \ref{s:Scalar}. Moreover, in the
perturbative terms by inhomogeneous perturbations of the phase-space
fields, there are additional corrections called ``counterterms'' which
are completely fixed by the consistency requirements. They can  be
understood as determining the dependence of $\bar{\alpha}$ on
inhomogeneities going beyond the background behavior which is more
straightforward to compute directly from expectation values of
inverse-triad operators. Some counterterms also contain additional
spatial derivatives compared to the classical terms, which can be
interpreted as contributions from a derivative expansion of non-local
inverse-triad effects, making use of surface integrations of the
densitized triad, or flux operators, in inhomogeneous settings.

\subsubsection{Spherical Symmetry}
\label{s:InvSph}

A second class of models in which inverse-triad corrections have been included
consistently, this time non-perturbatively, is spherically symmetric
models. Several different cases have been investigated: Poisson Sigma Models
\cite{SphSymmPSM} (see \cite{Ikeda,IkedaIzawa,PSM,Strobl} for the classical
models) and different versions of Lema\^{i}tre--Tolman--Bondi models
\cite{LTB,LTBII}. In these models, it is noteworthy that non-trivial quantum
corrections are possible even without any deformation of the constraint
algebra, a property which we will discuss in more detail later. 

We quote the corrected constraints in terms of triad variables rather than the
metric $g_{ij}$ because one of the triad components is directly responsible
for the corrections. (In the full theory, by comparison, it is primarily $\det
g$ which gives rise to inverse-triad corrections. Because $\det g$ equals the
squared determinant of the densitized triad, in the general case it will make
no difference what variables we use.)  As spherically symmetric phase space
variables, with radial coordinate $x$ (not necessarily the area radius) and
azimuth angle $\varphi$, we then have the radial component $E^x$ and angular
component $E^{\varphi}$ of the densitized triad together with the radial
component $K_x$ and angular component $K_{\varphi}$ of extrinsic curvature
\cite{SphSymm,SphSymmHam}. The metric is related to $E^x$ and $E^{\varphi}$ by
$g_{xx}= (E^{\varphi})^2/|E^x|$ and $g_{\varphi\varphi}=
|E^x|\sin\vartheta$. Consistent deformations of the Hamiltonian constraint
(with unmodified diffeomorphism constraint) have the form
\begin{align}
H^Q_{\rm grav}[N]=-\frac{1}{2G}\int \md x\,
N\bigg[&\alpha|E^x|^{-\frac{1}{2}} K_{\varphi}^2E^{\varphi}+2\bar{\alpha}
K_{\varphi}K_x|E^x|^{\frac{1}{2}}+ \notag\\ 
&+\alpha_{\Gamma}|E^x|^{-\frac{1}{2}}(1-\Gamma_{\varphi}^2)E^{\varphi}+
2\bar{\alpha}_{\Gamma}\Gamma_{\varphi}'|E^x|^\frac{1}{2} \bigg] \label{myEffectiveHamiltonian}
\end{align}
with correction functions $\alpha$, $\bar{\alpha}$,
$\alpha_{\Gamma}$ and $\bar{\alpha}_{\Gamma}$. In the second line,
$\Gamma_{\varphi}= -(E^x)'/2E^{\varphi}$ is the angular component of
the spin connection.

The four correction functions are not independent but must satisfy
\cite{LTBII}
\begin{equation} \label{alphaGamma}
 (\bar{\alpha}\alpha_{\Gamma}-\alpha\bar{\alpha}_{\Gamma}) (E^x)'
 + 2(\bar{\alpha}'\bar{\alpha}_{\Gamma}-
 \bar{\alpha}\bar{\alpha}_{\Gamma}') E^x=0
\end{equation}
for the Poisson bracket of two Hamiltonian constraints to be anomaly-free.  If
the two terms in this equation vanish separately, a case studied in
\cite{LTBII}, they imply $\alpha_{\Gamma}\propto\alpha$ and
$\bar{\alpha}_{\Gamma}\propto \bar{\alpha}$ for a closed constraint
algebra. For correction functions defined such that they approach the
classical value one for large arguments, $\alpha_{\Gamma}= \alpha$ and
$\bar{\alpha}_{\Gamma}= \bar{\alpha}$.

From the Poisson bracket $\{H[N],D[N^x]\}$, the only restriction is that both
correction functions depend only on the radial triad component $E^x$, not on
$E^{\varphi}$. (This fact is easily understandable from transformation
properties of the components: $E^x$ is a scalar in the radial manifold while
$E^{\varphi}$ is a scalar density \cite{SphSymm}.)  Only the functions
$\bar{\alpha}$ and $\bar{\alpha}_{\Gamma}$ appear in the deformed constraint
algebra (\ref{HH}) via $\beta=\bar{\alpha}\bar{\alpha}_{\Gamma}$, not $\alpha$
or $\alpha_{\Gamma}$. By changing only $\alpha$ and $\alpha_{\Gamma}$, one can
modify the spherically symmetric constraints while keeping their derivative
order and the constraint algebra unchanged: In spherical symmetry, the
classical dynamics does not follow uniquely from the hypersurface-deformation
algebra.

\subsubsection{$2+1$ gravity}

In spherically symmetric models and for perturbative inhomogeneities around
isotropic models, consistent deformations of the hypersurface-deformation
algebra have been found by computing Poisson brackets of effective
constraints, obtained by amending the classical constraints by correction
functions. In $2+1$-dimensional models, there are detailed calculations
\cite{TV} of partially off-shell constraint algebras even at an operator
level. The results confirm the general form of consistent deformations seen
with effective constraints.

\subsection{Holonomy corrections}
\label{s:Hol}

In spherically symmetric models, also a limited version of holonomy
corrections has been implemented consistently, those that involve only
the scalar component $K_{\varphi}$ of extrinsic curvature but not the
component $K_x$ of density weight one \cite{JR}. One can therefore
consistently substitute $-i\delta^{-1}\exp(i\delta K_{\varphi})$ for
$K_{\varphi}$, but there is no known consistent way to use $\exp(i\int_e
K_x{\rm d}x)$ for $K_x$.  Accordingly, no spatial integration or
discretization is required to ensure the existing forms of consistent
correction functions to be scalar and to keep the $\{H,D\}$-part of
the constraint algebra unmodified. This type of correction thus does
not lead to non-locality, as holonomy corrections usually do owing to
the spatial integrations involved in their definition. In this case,
the form of the deformation is similar to the one obtained for
inverse-triad corrections, with a Poisson bracket (\ref{HH}) for a
correction function now depending on extrinsic curvature instead of
the densitized triad.

If we parameterize the Hamiltonian constraint as
\begin{align}
H^Q_{\rm grav}[N]=-\frac{1}{2G}\int \md x\,
N\bigg[&\alpha|E^x|^{-\frac{1}{2}} E^{\varphi}f_1(K_{\varphi},K_x)+
2\bar{\alpha} |E^x|^{\frac{1}{2}}f_2(K_{\varphi},K_x) \notag\\ 
&+\alpha_{\Gamma}|E^x|^{-\frac{1}{2}}(1-\Gamma_{\varphi}^2)E^{\varphi}+
2\bar{\alpha}_{\Gamma}\Gamma_{\varphi}'|E^x|^\frac{1}{2} \bigg]\,,
\end{align}
including inverse-triad corrections as well as holonomy corrections via two
new functions $f_1$ and $f_2$, anomaly freedom can be realized if $f_1=F_1^2$
and $f_2=K_x F_2$ provided that $F_2= F_1(\partial F_1/\partial K_{\varphi})
\alpha/\alpha_{\Gamma}$ \cite{JR}. If $F_1$ is independent of $E^x$, or at
least depends on this triad variable in a way different from inverse-triad
corrections, we obtain that $\alpha_{\Gamma}=\alpha$ and also
$\bar{\alpha}_{\Gamma}=\bar{\alpha}$ must be realized, restricting the set of
solutions of (\ref{alphaGamma}). Combinations of different corrections
therefore can reduce the freedom of choices seen for just a single type. If we
take $F_1=(\delta\gamma)^{-1} \sin(\gamma\delta K_{\varphi})$, as often done
for holonomy corrections, we see that
$F_2=(2\gamma\delta)^{-1}\sin(2\gamma\delta K_{\varphi})$. The algebraic
deformation is then given by
$\beta(E^x,K_{\varphi})=\bar{\alpha}\bar{\alpha}_{\Gamma} \partial
F_2/\partial K_{\varphi}$. For the example provided, this means
$\beta(K_{\varphi})= \cos(2\delta K_{\varphi})$ if we include only holonomy
corrections. Note that $\beta$ can be negative for holonomy corrections,
unlike for inverse-triad corrections.

A more general case of holonomy corrections, including even discretization and
non-locality, has been implemented consistently in $2+1$-dimensional gravity
with a non-vanishing cosmological constant \cite{ThreeDeform}. (A vanishing
cosmological constant in $2+1$ dimensions does not require deformations of the
constraint algebra, which is much simpler in this case.) As with inverse-triad
corrections in $2+1$ dimensions, also these calculations have been performed
at an operator level. Similarly to the spherically symmetric case, the
correction function is given by the trace of a holonomy used to write the
Hamiltonian constraint in loop variables.

For some perturbative models around Friedmann--Robertson--Walker backgrounds,
holonomy corrections have been included consistently, too. This is the case
for tensor \cite{tensor}, vector \cite{VectorHol}, and scalar modes
\cite{ScalarHol}. A new feature in \cite{VectorHol}, which did not show up in
any other case of consistent deformations of (\ref{Algebra}) so far, is that
the Poisson bracket $\{H,D\}$ could be consistently modified (even if $D$
itself remains classical). However, this possibility has been ruled out by
more restricted consistent deformations of scalar modes \cite{ScalarHol}. Also
here, the correction function is of the form $\cos(2\delta c)$ with the
isotropic connection component $c$. It is similar to the correction function
for holonomy corrections in spherical symmetry, and also becomes negative for
large curvatures, with $\delta c\sim\pi/2$. Implications will be discussed
later. As in spherical symmetry, no non-locality effects have yet been
implemented for holonomy corrections in nearly isotropic cosmology.

\subsection{Discretization}

Effective constraints of loop quantum gravity in inhomogeneous situations
naturally include discretization (or a derivative expansion of spatially
discretized terms) because the basic variables, holonomies and fluxes, are
defined as spatial integrations of non-scalar quantities. Also spatial
derivatives in Hamiltonians must be replaced suitably by finite
differences. Modelling classical constraints with these variables to ensure
the correct classical limit of the resulting theory then requires one to refer
to the field values at different points even for classically local
expressions.  For this type of corrections, independently of holonomy
corrections, no consistently deformed algebra has been formulated explicitly,
but work on consistent discretizations exists
\cite{UniformDisc,PerfectAction,BrokenAction} and indicates that deformations
should occur also here.

\section{Hypersurface deformations}

The meaning of the hypersurface-deformation algebra has been discussed in
detail in the classic references \cite{LagrangianRegained} and
\cite{Regained}. Nevertheless, it is useful to go through some of
the arguments once again with a fresh perspective suggested by the deformed
algebras found recently and summarized in the preceding section.

\subsection{Spatial diffeomorphisms}

Most (but not all) deformations found so far in loop quantum gravity
leave the spatial part of the hypersurface-deformation algebra intact,
which will also be one of our assumptions in this article. There are
several reasons for this assumption: First, spatial diffeomorphisms
can be implemented directly in loop quantum gravity by moving graphs
in the spatial manifold used to set up the canonical formulation. This
action is the same as the one on classical fields, and so one would
not expect corrections to the diffeomorphism constraint at an
effective level. If one just assumes that the part of the constraint
algebra associated with a vector field $\delta N^i$ generates
relabellings $x^i\mapsto x^i+\delta N^i$ of points in the spatial
manifold, any field on space must automatically change by the Lie
derivative along $\delta N^i$. Since spaces in a very general sense
are described mathematically by labelling their elements in some way,
while physics should be insensitive to how the labels are chosen, it
is natural to expect a relabelling symmetry to be present at an
effective level, even if the fundamental spatial structure may become
discrete or non-commutative. From the relation $\{F,D[\delta N^i]\}=
{\cal L}_{\delta N^i}F$ and the usual expressions for Lie derivatives
of the fundamental fields, one can then uniquely derive the
phase-space expression that the diffeomorphism constraint must take
\cite{LagrangianRegained}. In particular, it is always linear in the
momenta of the fields, a consequence which we will make use of later
on. For the fields considered here, this implies $D_{\rm scalar}[N^i]=
\int{\rm d}^3x N^ip_\phi \phi_{|i}$ for a scalar field and
$D_{\rm grav}[N^i]= -2\int{\rm d}^3x N^i \pi_i{}^j_{|j}$ for gravity in
ADM variables.

Once the diffeomorphism constraint is determined, it must obviously satisfy
the spatial part (\ref{Algebra}) of the classical constraint algebra as well
as (\ref{HD}), as long as the corrected $H_{(\beta)}[N]$ remains a scalar. The
latter assumption (that $H_{(\beta)}[N]$ be a spatial scalar) appears safe,
too, because of the nature of effective constraints as integrated functionals
on a spatial manifold. In what follows, we will make use not only of the
assumption that the spatial part of the hypersurface-deformation algebra
remains unmodified, but also of several further consequences regarding the
form of the diffeomorphism constraint. Most importantly, the diffeomorphism
constraint appears on the right-hand side of (\ref{HH}); thus, the expression
it takes will influence the Hamiltonian constraint determined from the
constraint algebra.

\subsection{Transversal deformations}
\label{s:Trans}

The modification by $\beta$ in (\ref{HH}) occurs for the commutator of two
transversal deformations of spatial hypersurfaces along their normal vectors,
by two different and position-dependent amounts $N_1$ and $N_2$. This part of
the deformation algebra is distinguished from the spatial part not only in
that it is of dynamical content, owing to the presence of the Hamiltonian
constraint and matter Hamiltonians. Also, the use of the normal vector to point
the deformation normally implies a dependence on the space-time metric
$\stmet_{\mu\nu}$, containing phase-space degrees of freedom. The algebra, as
a consequence, acquires structure functions rather than just structure
constants as suffice for the part of spatial deformations. Implications of
structure functions for canonical quantization, mainly negative ones owing to
additional difficulties in commutator relationships, are well known; in the
present context they are, perhaps more positively so thanks to interesting
implications, realized as a general source of possible deformations by quantum
corrections.

Unlike the spatial part of the deformation algebra, which directly shows its
relation to infinitesimal deformations by the presence of the Lie derivative,
relating the $\{H,H\}$ part of the algebra to transversal deformations is not
so obvious. As indicated by the algebra, we consider two transversal
deformations by lapse functions $N_1$ and $N_2$, done in a row but in the two
different possible orders. Starting with an initial hypersurface $S_{\rm in}$,
we obtain two intermediate ones, $S_1$ by deforming $S_{\rm in}$ by $N_1$
along the normal and $S_2$ by deforming $S_{\rm in}$ by $N_2$ along the
normal. From those two intermediate hypersurfaces, we obtain two final
hypersurfaces, $S_{\rm fin}^{(1)}$ by deforming $S_1$ by $N_2$ along the new
normal of $S_1$ and $S_{\rm fin}^{(2)}$ by deforming $S_2$ by $N_1$ along the
new normal of $S_2$. Comparing the two final hypersurfaces should then yield a
commutator of deformations according to (\ref{HHclass}). In the process of
computing the normals of $S_{\rm in}$, $S_1$ and $S_2$, the metric tensor must
be used. We will not fix the signature $\sigma=\pm 1$ of the metric for our
calculations in order to be able to incorporate possible sign changes due to
quantum corrections, as suggested by holonomy corrections where $\beta$ can
turn negative. (For Lorentzian signature with $\sigma=-1$, we choose the time
part of the metric to carry the minus sign.)

For simplicity, and without loss of generality, we choose space-time
coordinates such that $S_{\rm in}$ is given by a constant-time slice, $S_{\rm
  in}\colon y^i\mapsto (t_{\rm in},y^i)$ with some spatial embedding
coordinates $y^i$. The general expression for the future-pointing unit normal
to a hypersurface $y^i\mapsto x^{\mu}(y^i)$,
\begin{equation} \label{Normal}
 n^{\mu}= \sigma\frac{\stmet^{\mu\mu'}
   \epsilon_{\mu'\nu\lambda\kappa} \partial_{y^1}x^{\nu} \partial_{y^2}x^{\lambda} \partial_{y^3}x^{\kappa}}{||
   \cdot||}
\end{equation}
(with $||\cdot||$ denoting the norm of the numerator) then simplifies to
$n_{\rm in}^{\mu}= \sigma\stmet^{\mu 0}/\sqrt{|\stmet^{00}|}$.

The intermediate hypersurfaces, with infinitesimal transversal deformations,
are obtained as 
\begin{eqnarray*}
 S_1\colon y^i\mapsto x^{\mu}(y^i)+ N_1(y^i) n_{\rm
  in}^{\mu}&=& (t_{\rm in},y^i)+\sigma N_1(y^i) \stmet^{\mu 0}(y^i)/\sqrt{|\stmet^{00}|} \\
 S_2\colon y^i\mapsto x^{\mu}(y^i)+ N_2(y^i) n_{\rm
  in}^{\mu}&=& (t_{\rm in},y^i)+\sigma N_2(y^i) \stmet^{\mu 0}(y^i)/\sqrt{|\stmet^{00}|}\,.
\end{eqnarray*}
From these expressions, we obtain the new normals by the general formula
(\ref{Normal}), expanded to first order in the lapse functions for
infinitesimal deformations:
\begin{eqnarray*}
 n_1^{\mu} &=& \sigma\frac{\stmet^{\mu 0}}{\sqrt{|\stmet^{00}|}}+
       \left(-\sigma \stmet^{\mu
       i}+ \frac{\stmet^{\mu 0}\stmet^{i0}}{|\stmet^{00}|}\right) 
\partial_i N_1+N_1 X+ O(N_1^2)\\
 &=& \sigma\frac{\stmet^{\mu 0}}{\sqrt{|\stmet^{00}|}}-\sigma g^{\mu
    i} \partial_i N_1 +N_1 X+O(N_1^2)\\
 n_2^{\mu} &=& \sigma\frac{\stmet^{\mu 0}}{\sqrt{|\stmet^{00}|}}+
       \left(-\sigma \stmet^{\mu
       i}+ \frac{\stmet^{\mu 0}\stmet^{i0}}{|\stmet^{00}|}\right) 
\partial_i N_2+N_2 X+ O(N_2^2)\\
 &=& \sigma\frac{\stmet^{\mu 0}}{\sqrt{|\stmet^{00}|}}-\sigma g^{\mu
    i} \partial_i N_2+N_2 X+O(N_2^2)
\end{eqnarray*}
with the spatial metric $g^{\mu\nu}= \stmet^{\mu\nu}-\sigma n_{\rm
  in}^{\mu} n_{\rm in}^{\nu}$ on the initial slice. The coefficient
  $X$ denotes a combination of metric components and their
  derivatives; the precise form will not be important because these
  terms, depending on $N_1$ and $N_2$ but not on their derivatives,
  will drop out of the final commutator results.  The two final
  hypersurfaces are then parameterized as
\begin{eqnarray*}
 S_{\rm fin}^{(1)}\colon y^i&\mapsto& x^{\mu}(y^i)+ N_1(y^i) n_{\rm in}^{\mu}+
 N_2(y^i) n_1^{\mu}\\
  &=& (t_{\rm in},y^i)+\sigma N_1(y^i) \frac{\stmet^{\mu
 0}}{\sqrt{|\stmet^{00}|}}+\sigma N_2(y^i) \left(\frac{\stmet^{\mu
   0}}{\sqrt{|\stmet^{00}|}}- g^{\mu i} \partial_i N_1(y^i)\right)+ 
N_1N_2X+O(N_1^2)\\
S_{\rm fin}^{(2)}\colon y^i&\mapsto& x^{\mu}(y^i)+ N_2(y^i) n_{\rm in}^{\mu}+
 N_1(y^i) n_2^{\mu}\\
  &=& (t_{\rm in},y^i)+\sigma N_2(y^i) \frac{\stmet^{\mu
 0}}{\sqrt{|\stmet^{00}|}}+\sigma 
 N_1(y^i) \left(\frac{\stmet^{\mu
   0}}{\sqrt{|\stmet^{00}|}}- g^{\mu i} \partial_i N_2(y^i)\right)
 +N_2N_1X+O(N_2^2)\,.
\end{eqnarray*}
With these expressions it is easy to notice that, writing $S_{\rm
  fin}^{(1)}\colon y^i\mapsto x_{{\rm fin},1}^{\mu}(y^i)$, we have
\[
 S_{\rm fin}^{(2)} \colon y^i\mapsto x_{{\rm fin},2}^{\mu}(y^i)= x_{{\rm
     fin},1}^{\mu}(y^i)+ \delta S^{\mu}(y^i)
\]
with
\begin{equation} \label{deltaS}
 \delta S^{\mu}(y^i)= -\sigma 
 g^{\mu i} (N_1\partial_i N_2- N_2\partial_i N_1)\,.
\end{equation}
To leading order in the lapse functions, $\delta S^{\mu}(y^i)$ (depending only
on spatial metric components $g^{\mu i}$) is orthogonal to the normal vector
and thus amounts to an infinitesimal spatial diffeomorphism along the
hypersurface. The spatial deformation $\delta S^{\mu}$ in (\ref{deltaS}) is
obtained from the commutator of two normal deformations, and it reproduces the
normal part of the algebra (\ref{HHclass}) for $\sigma=-1$. A change of sign
in the structure function is equivalent to signature change. (Formally, this
implication of signature change can also be seen by replacing $N$ with $iN$.)

So far we have assumed the classical manifold structure and geometry
in order to compute the normal vectors. The deformed algebra
(\ref{HH}) can be achieved formally by using $\beta g^{\mu\nu}$
instead of the inverse metric $g^{\mu\nu}$. For inverse-triad
corrections, such a modification would be expected because it affects
all inverse components of the metric in Hamiltonians. Nevertheless,
the appearance of the correction function in the constraint algebra
must have a more general origin than just modifying any appearance of
the inverse metric because a deformation of the same form is obtained
for some versions of holonomy corrections. The latter do not affect
inverse-metric components but rather appearances of extrinsic
curvature or the Ashtekar--Barbero connection. However, only the
spatial metric appears in the structure functions of the constraint
algebra; deformations, therefore, cannot be reduced to simply applying
the usual corrections of loop quantum gravity to the structure
functions. Such a procedure would be questionable, anyway, because the
structure functions are not quantized but rather arise from the
algebra satisfied by effective quantum constraints, with corrections
following in a more indirect way.

\section{Constraints and space-time structure}

Quantum-geometry corrections change the hypersurface-deformation algebra and
accordingly the space-time structure: Normal deformations of spatial slices
then behave differently from the classical case.  Corresponding actions cannot
be covariant in the usual sense, but they are still covariant in a deformed
sense, under an algebra of the type (\ref{HH}). In the absence of a
corresponding space-time tensor calculus, it is difficult to imagine the form
of actions covariant with respect to the new space-time structures. But
fortunately, such actions can be systematically derived from the constraint
algebra, or regained in the language of \cite{LagrangianRegained,Regained}.

In this and the following section we will go in some detail through
the steps outlined in \cite{LagrangianRegained}, focusing our
discussion on those that use assumptions no longer valid if the
classical space-time structure cannot be taken for granted. According
to the form of the deformed constraint algebra used here, and as a
rather general consequence of canonical quantum gravity, the spatial
structure on the one hand and the structure of hypersurface
deformations within space-time, on the other, will play rather
different roles. The algebraic effects considered here are thus truly
dynamical and do not arise at the kinematical level of spatial
manifolds.

\subsection{Locality}
\label{s:Local}

Once the spatial structure is fixed, the next object to consider is the change
of the spatial metric under a normal deformation of a spatial
slice. Classically, this deformation is given by the extrinsic-curvature
tensor, $\{\met{x}, H[\delta N]\}= 2K_{ij}(x) \delta N$, and it plays an
important role in \cite{LagrangianRegained} in helping to show that the
Hamiltonian constraint must be a local expression in the momentum:
Identifying
\begin{equation}
\frac{\delta H[\delta N]}{\delta\mom{x}}= \{\met{x}, H[\delta N]\}= 
2K_{ij}(x)\delta N(x)
\end{equation}
implies that $H[\delta N]$ must be local in the momentum $\mom{x}$ without any
dependence on $\pi^{ij}$-derivatives. The specific form of $K_{ij}$ as
extrinsic curvature does not matter for this conclusion, but it is important
that it is a local function, and that no derivatives of $\delta N$ appear on
the right-hand side.

In the presence of deformed space-time structures, we cannot safely assume
that transversal metric deformations are given in terms of extrinsic
curvature. For the explicit examples of deformed constraint algebras, it is
known that the relationship between momentum variables and extrinsic curvature
deviates from the classical one; see e.g.\ the discussion in \cite{LTBII}. It
should then be possible for the change of the metric under a transversal
deformation, while still being related to the momentum of the metric, to have
a modified relationship with extrinsic curvature. In the absence of a
geometrical interpretation of the change of the metric, one can compute it
only by using the canonical formula $\{g_{ij}(x), H[\delta N]\}$; but then,
one piece of independent information is lost and we cannot derive locality
properties of the Hamiltonian constraint. If $H[\delta N]$ is local in the
momentum, $\{g_{ij}, H[\delta N]\}$ is local and vice versa, but there is no
independent general statement that could determine whether locality is
realized.

Instead, we will make use of the following line of arguments: We know that the
classical constraint must be local without spatial derivatives of $\pi^{ij}$,
and in most cases the form of corrections expected from loop quantum gravity
tells us what locality properties new terms have. Most of them are indeed
non-local, for instance those arising from the use of holonomies as
exponentiated line integrals of a connection related to extrinsic curvature,
or inverse-triad corrections depending on fluxes through extended surfaces. In
derivative expansions, whole series of spatial derivatives of $\pi^{ij}$ or
$g_{ij}$ will result. The form of the corrections and their impact on
effective constraints can thus be used to decide whether local or non-local
constraints should be expected. The arguments put forward to regain the form
of the constraint will then have to be adapted, depending on the locality
properties realized. In most cases, effective equations include a derivative
expansion, approximating non-local features locally. We can then assume a
local Hamiltonian constraint, but, in contrast to the classical case, must
take into account additional derivatives, for instance of $K_{ij}$.

Similar considerations can be applied to the question of whether the matter
Hamiltonian must be local in the momentum. Here, the assumptions made in
\cite{LagrangianRegained} appear safer in the context of deformed algebras
than those for the corresponding gravitational terms. Instead of looking at
transversal deformations of the spatial metric, we look at transversal
deformations of the matter field, assumed to be a scalar to be specific. The
relation $\{\phi(x),H[\delta N]\}= V(x)\delta N$ then replaces the
gravitational relation involving extrinsic curvature, with $V(x)$ interpreted
as the velocity of the scalar field. In contrast to the gravitational part,
there are some quantum corrections in matter Hamiltonians that, while changing
the specific expression for $V(x)$, leave its local nature
intact.\footnote{These local modifications are ``point holonomies''
  \cite{FermionHiggs} obtained by exponentiating the scalar pointwise, without
  integrations as in holonomies. Of similar nature are the modifications of
  the scalar sector proposed in \cite{ExpScalar}, replacing the scalar
  momentum with bounded exponentials. No constraint analysis has been made for
  the latter case coupled to gravity, but from our results in Section
  \ref{s:Scalar} it will become clear that it cannot correspond to an
  undeformed algebra.} Thus, in some cases we can assume the matter
Hamiltonian to be local in the momentum even in the presence of corrections
making the gravitational part non-local without a derivative expansion. This
difference between gravitational and matter Hamiltonians may play an important
role for the interplay of different contributions to the constraints ensuring
that the algebra closes.

\subsection{Gravity and matter}

There is a useful argument showing that the gravity and matter parts of the
constraints $D[N^i]= D_{\rm grav}[N^i]+ D_{\rm matter}[N^i]$ and $H[N]= H_{\rm
  grav}[N]+ H_{\rm matter}[N]$ must satisfy the hypersurface-deformation
algebra separately, provided that matter Hamiltonians do not depend on the
gravitational momentum $\mom{x}$ and the gravitational constraint does not
contain spatial derivatives of $\mom{x}$. In this case,
\begin{equation} \label{HHmatter}
 \{H[N_1],H[N_2]\}= \{H_{\rm grav}[N_1],H_{\rm grav}[N_2]\}+ \{H_{\rm
   matter}[N_1],H_{\rm matter}[N_2]\}\,.
\end{equation}
The assumption is realized classically for a scalar field, for instance, and
so one can consider its simpler algebraic regaining procedure independently of
the gravitational part. With quantum corrections, however, the assumption can
be violated easily, depending on the type of the correction. Matter fields are
usually introduced in loop quantum gravity via the values they take at the
vertices of a spin network. Spatial derivatives as they occur in the
Hamiltonians must be discretized and replaced by finite differences of the
values at neighboring vertices before they can be quantized. Depending on how
the differencing is done, one may have to refer to the gravitational
connection, making the matter constraint dependent on the gravitational
momentum. Another source of such a dependence may be counterterms as
introduced in \cite{ConstraintAlgebra}, required to close the constraint
algebra. An extra momentum dependence can be avoided for a scalar field, but
there may be reasons to prefer more complicated quantizations.

Coming back to the results found in the preceding subsection on locality, we
can see a potential obstruction to the existence of consistent deformations of
the classical constraint algebra. There are corrections expected from loop
quantum gravity, most notably holonomy corrections, which are non-local in the
connection and thus make the gravitational part of the Hamiltonian constraint
non-local in the gravitational momentum. A scalar Hamiltonian in the presence
of the same corrections, on the other hand, remains local in its momentum. If
the gravitational part and the matter part are to satisfy the same deformed
algebra for consistency, the mismatch of locality properties could be seen as
an obstacle to the existence of a consistent deformation: The function $\beta$
of (\ref{HH}) would be non-local in one contribution and local in another one,
preventing one from adding up the constraint contributions to a consistent
whole. However, the situation is not obviously inconsistent because the same
property giving rise to the mismatch, non-locality and the presence of
derivatives of $\pi^{ij}$, also violates the assumptions that led one to
conclude that gravity and matter satisfy the hypersurface-deformation algebra
independently. Non-locality, in a derivative expansion of holonomy corrections
in effective constraints, makes the gravitational constraint depend on spatial
derivatives of the momentum $\mom{x}$, such that cross-terms between gravity
and matter in (\ref{HHmatter}) no longer cancel.  It is reassuring that
properties of non-locality thus restore the a-priori possibility of
consistency, but the necessary appearance of gravity-matter cross-terms makes
the explicit construction of consistent deformations for non-local
momentum-dependent corrections more difficult than for local ones. As recalled
in Sec.~\ref{s:Hol}, results in spherical symmetry are indeed much easier to
find in local versions of the corrections. Also for perturbative
inhomogeneities as in \cite{ScalarHol} one so far assumes a local, pointwise
form of holonomy corrections. The manipulations required for non-local
modifications to be consistent appear to be rather complicated, a fact which
may explain the difficulties found in constructing consistent deformations
corresponding to the non-local holonomy or discreteness corrections. (On the
other hand, tying matter terms more closely to gravitational ones rather than
having them algebraically separated as in (\ref{HHmatter}) may be of interest
in the context of unification.)

\section{Algebraically regaining Hamiltonians}

With these preparatory discussions, we can now begin to enter details
of regaining Hamiltonians from deformed constraint algebras. There are
several interesting applications and generalizations of the methods of
\cite{LagrangianRegained}, which we develop in different cases.

\subsection{Spherical symmetry}

Before looking at the general theory, it is instructive to specialize
the calculations to spherical symmetry. Some steps will simplify, and
it will be interesting to compare the differences in uniqueness for
different degrees of symmetry. As already noted in
Sec.~\ref{s:InvSph}, in spherical symmetry the classical dynamics does
not follow uniquely from the algebra.

For the sake of easier comparison with calculations of modified constraints
motivated by loop quantum gravity, we will present equations in this
subsection for triad variables.  A spherically symmetric spatial densitized
triad has two components $E^x$ and $E^{\varphi}$, for the radial coordinate
$x$ and one angular coordinate $\varphi$, which determine the spatial metric
by $g_{xx}=(E^{\varphi})^2/|E^x|$ and $g_{\varphi\varphi}=\sin^2\vartheta
|E^x|$. We will assume $E^x>0$ to avoid some sign factors.

Instead of working with spatial curvature tensors, in this context it turns
out to be useful to refer to the angular spin-connection component and its
spatial and functional derivatives,
\begin{align}
\gp =& -\frac{\erp}{2\ep} \quad,\quad \gp' = -\frac{(E^{x})''}{2\ep} + \frac{\erp\epp}{2(\ep)^2}\\
\fund{\gp(y)}{E^x(x)} =& -\frac{1}{2\ep(y)}\dyxp \quad,\quad
\fund{\gp(y)}{\ep(x)} = \frac{\erp(y)}{2\ep(y)^2}\dyx \label{GP1}\\
\fund{\gp'(y)}{E^x(x)} =& -\frac{1}{2\ep(y)}\dyxpp + \frac{\epp(y)}{2\ep(y)^2}\dyxp\\
\fund{\gp'(y)}{\ep(x)} =& \frac{(E^{x})''(y)}{2\ep(y)^2}\dyx + \frac{\erp(y)}{2\ep(y)^2}\dyxp
 - \frac{\erp(y)\epp(y)}{\ep(y)^3}\dyx\,. \label{GP3}
\end{align}
(The radial component of the spin connection does not have any gauge-invariant
contribution \cite{SphSymm}.)

Momenta of the densitized triad are classically given by
extrinsic-curvature components $K_x$ and $K_{\varphi}$ with
$\{K_x(x),E^x(y)\}= 2G\delta(x,y)$ and
$\{K_{\varphi}(x),E^{\varphi}(y)\}= G\delta(x,y)$. With these
properties, the commutator relationship (\ref{HH}) to exploit here reads
\begin{align}
\{H(x),H(y)\} ={}& G\int \dif^3 z \left(2\fund{H(x)}{\kx(z)}\fund{H(y)}{E^x(z)} - 2\fund{H(y)}{\kx(z)}\fund{H(x)}{E^x(z)}\right.\notag\\
 &\left.{\hspace{0.7cm}} + \fund{H(x)}{\kp(z)}\fund{H(y)}{\ep(z)} - \fund{H(y)}{\kp(z)}\fund{H(x)}{\ep(z)}\right)\notag\\
={}& \beta(x)\frac{E^x(x)}{\ep(x)^2}D(x)\dxyp - (x\leftrightarrow y)
\end{align}
with the local diffeomorphism constraint 
\begin{equation}
 D(x)= \frac{1}{2G} (2E^{\varphi}K_{\varphi}'- K_x(E^x)')\,.
\end{equation}
With a modified Hamiltonian, $K_x$ and $K_{\varphi}$ may no longer be
components of extrinsic curvature. However, they are still canonically
conjugate to $E^x$ and $E^{\varphi}$, and we continue to use the same
letters for momentum variables.

For now, we will be looking only for constraints with quadratic ``kinetic''
term in momenta and no non-locality or spatial derivatives of $K$,
\be  \label{Hsecond}
 H = \jig{00} +  \jig{11}\kx\kp + \jig{20}\kx\kx + \jig{02}\kp\kp
\ee 
(without linear terms, assuming time reversal symmetry),
and have linear functional derivatives
\begin{align}
G\fund{H(x)}{\kx(z)} &= \left(A_{1}(x)\kx(x) + B_{1}(x)\kp(x)\right)\dxz\\
G\fund{H(x)}{\kp(z)} &= \left(A_{2}(x)\kx(x) + B_{2}(x)\kp(x)\right)\dxz\,.
\end{align}
We then identify $G\jig{11} = A_{2}=B_1$, $G\jig{02} = B_{2}/2$, $G\jig{20} =
A_{1}/2$, which may all depend on the triad components.  The Poisson bracket
of two Hamiltonian constraints becomes
\begin{align}
\{H(x),H(y)\} ={}& \fund{H(y)}{E^x(x)}\left(2A_{1}\kx(x) + 2B_{1}\kp(x)\right)\notag\\
&{}+ \fund{H(y)}{\ep(x)}\left(A_{2}\kx(x) + B_{2}\kp(x)\right) -
(x\leftrightarrow y) \notag\\
={}& 
\beta(x)\frac{E^x(x)}{G\ep(x)^2}\left(\ep(x)\kp'(x) -
  \frac{1}{2}\kx(x)\erp(x)\right)\dxyp - (x\leftrightarrow y) \,. \label{allorder}
\end{align}
We evaluate its implications by comparing coefficients of $K_x$
and $K_{\varphi}$. In this section, we will assume that $\beta$ does
not depend on $K_x$ or $K_{\varphi}$, thus considering the case of
inverse-triad corrections.

For $\kx=0$, $\kp=0$, the equation is automatically satisfied.  For the
first-order coefficients in \kx, we operate with $\delta/\delta K_x$ and
then set $\kx=0$, $\kp=0$:
\begin{equation}
\begin{split}
\left(2\fund{\jig{00}(y)}{E^x(x)}A_{1}(x) \right.&+\left. \fund{\jig{00}(y)}{\ep(x)}A_{2}(x)\right)\dxz - (x\leftrightarrow y) \\
&{}= -
\frac{\beta E^x(x)\erp(x)}{2G\ep(x)^{2}}\dxyp\dxz - (x\leftrightarrow y) \,.
\label{FirstKxSph}
\end{split}
\end{equation}
For functional derivatives of $\jig{00}$ by $E^x$ and $E^{\varphi}$, we must
know the general triad-dependent terms possible. In addition to a direct
dependence on the fields, $\jig{00}$ can depend on the triad via spatial
curvature which, in turn, depends on the spin connection and its
derivatives. We thus have to expect a dependence on $E^x$, $E^{\varphi}$,
$\Gamma_{\varphi}$ and $\Gamma_{\varphi}'$. Higher derivatives are not
included because here, as in (\ref{Hsecond}), we expand only to second order
in momenta and derivatives.

We then have the chain rule
\be
\fund{\jig{00}(y)}{E^x(x)} = \pard{\jig{00}(y)}{\gp(y)}\fund{\gp(y)}{E^x(x)}
+ \pard{\jig{00}(y)}{\gp'(y)}\fund{\gp'(y)}{E^x(x)}
+ \pard{\jig{00}(y)}{E^x(y)}\fund{E^x(y)}{E^x(x)} 
\ee
and a similar relation for $\delta\jig{00}(y)/\delta E^{\varphi}(x)$
to rewrite (\ref{FirstKxSph}).
We substitute our expressions (\ref{GP1})--(\ref{GP3}) for $\delta
\Gamma_{\varphi}(y)/\delta E^x(x)$ and so on, multiply with test functions
$a(x)$, $b(y)$, and $c(z)$, and integrate over $x$, $y$, and $z$.  We state
the result obtained after several integrations by parts:
\begin{equation}
\begin{split}
\int &{}\dif y \bigg[ -(a' c A_{1})\frac{b}{\ep}\pard{\jig{00}}{\gp} + (a' c A_{1})\frac{b\epp}{(\ep)^{2}}\pard{\jig{00}}{\gp'} \\
&{}+ (a' c A_{2})\frac{b\erp}{2(\ep)^{2}}\pard{\jig{00}}{\gp'} + 2(a' c A_{1})b\bigg(\frac{1}{\ep}\pard{\jig{00}}{\gp'}\bigg)' \\
&{}+ (a'' c A_{1})\frac{b}{\ep}\pard{\jig{00}}{\gp'} - \frac{\beta E^x\erp}{2G(\ep)^{2}}a' c b \bigg] - (a \leftrightarrow b) = 0\,.
\end{split}
\end{equation}
(Several terms that cancel in the antisymmetrization with respect to
$a$ and $b$ have not been written explicitly.)  Collecting the
coefficients of $c(a''b - b''a)$ and $c(a'b-b'a)$, respectively, we
get 
\begin{eqnarray}
 \frac{A_{1}}{\ep}\pard{\jig{00}}{\gp'} &\!\!=\!\!& 0\,, \label{A1}  \\
-\frac{A_{1}}{\ep}\pard{\jig{00}}{\gp} + \frac{A_{1}\epp}{2(\ep)^{2}}\pard{\jig{00}}{\gp'} + \frac{A_{2}\erp}{2(\ep)^{2}}\pard{\jig{00}}{\gp'} 
+ \bigg(\frac{1}{\ep}\pard{\jig{00}}{\gp'}\bigg)'2A_{1} -
\frac{\beta E^x\erp}{2G(\ep)^{2}} &\!\!=\!\!& 0\,. \label{A1A2}
\end{eqnarray}

Going back to (\ref{allorder}) to look at the first order in \kp\
(and zeroth in \kx), and performing similar operations, we get
\begin{equation}
\begin{split}
\int &{}\dif y \bigg[ -(a' c B_{1})\frac{b}{\ep}\pard{\jig{00}}{\gp} + (a' c B_{1})\frac{b\epp}{(\ep)^{2}}\pard{\jig{00}}{\gp'} \\
&{}+ (a' c B_{2})\frac{b\erp}{2(\ep)^{2}}\pard{\jig{00}}{\gp'} + 2(a' c B_{1})b\bigg(\frac{1}{\ep}\pard{\jig{00}}{\gp'}\bigg)' \\
&{}+ (a'' c B_{1})\frac{b}{\ep}\pard{\jig{00}}{\gp'} - a'' b c
  \frac{\bar{\alpha}^{2}E^x}{\ep} - a' b c \bigg(\frac{\beta E^x}{G\ep}\bigg)' \bigg] - (a \leftrightarrow b) = 0\,.
\end{split}
\end{equation}
Collecting the coefficients of $c(a''b - b''a)$ and $c(a'b-b'a)$,
respectively, results in
\begin{eqnarray}
  \frac{B_{1}}{\ep}\pard{\jig{00}}{\gp'} -\frac{\beta E^x}{G\ep} &\!\!\!=\!\!\!&
  0 \label{B1}\\
-\frac{B_{1}}{\ep}\pard{\jig{00}}{\gp} + \frac{B_{1}\epp}{2(\ep)^{2}}\pard{\jig{00}}{\gp'} + \frac{B_{2}\erp}{2(\ep)^{2}}\pard{\jig{00}}{\gp'}
+ \bigg(\frac{1}{\ep}\pard{\jig{00}}{\gp'}\bigg)'2B_{1} -
 \bigg(\frac{\beta E^x}{G\ep}\bigg)' &\!\!\!=\!\!\!& 0\,.
\end{eqnarray}

Equation (\ref{B1}) implies
that $\delta \jig{00}/\delta \Gamma_{\varphi}'$ cannot be zero.  With
this condition, we find $A_1=0$ from (\ref{A1}),
\begin{equation} \label{A2B1}
A_{2} = \frac{\beta E^x}{G}
\bigg(\pard{\jig{00}}{\gp'}\bigg)^{-1} = B_{1}
\end{equation}
from (\ref{A1A2}) and (\ref{B1}), and
\be \label{H00diff}
-\frac{B_{1}}{\ep}\pard{\jig{00}}{\gp} +
\pard{\jig{00}}{\gp'}\bigg(\frac{B_{1}\epp}{(\ep)^{2}} +
\frac{B_{2}\erp}{2(\ep)^{2}}\bigg) +
2B_{1}\bigg(\frac{1}{\ep}\pard{\jig{00}}{\gp'}\bigg)' = \frac{1}{G}
\bigg(\frac{\beta E^x}{\ep}\bigg)' \,.
\ee
This tells us that
\be \label{B1B2}
G \frac{B_{1}}{\ep}\pard{\jig{00}}{\gp} =
\bigg(\frac{B_{2}}{B_{1}}\frac{E^x}{2\ep} + 1\bigg)\frac{\beta
  \erp}{\ep} + \frac{E^x}{\ep}\bigg(\beta' - 2 
\frac{B_{1}'}{B_{1}}\beta \bigg)\,.
\ee
To solve these equations, we introduce a function $b_1$ such that
$B_1=-\sqrt{|\beta|} b_1\sqrt{E^x}=A_2$. The factors are chosen so as to
cancel several terms in (\ref{B1B2}):
\[
 \frac{\beta (E^x)'}{E^{\varphi}}+ \frac{E^x}{E^{\varphi}} 
\left(\beta'-2\beta B_1'/B_1\right)=
 -2\beta\frac{E^x}{E^{\varphi}} \frac{b_1'}{b_1}\,.
\]

For the correct density weights in the first term in (\ref{B1B2}),
$B_2$ must be proportional to $E^{\varphi}$. (The other factors $B_1$
and $E^x$ are scalar and cannot change the density weight.) With
another free function $b_2$, we write $B_2=-b_1 b_2\sqrt{|\beta|}
E^{\varphi}/\sqrt{E^x}$, with factors other than $E^{\varphi}$ chosen
for later convenience.  The coefficients $A_1$, $A_2$, $B_1$ and $B_2$
determine the form of momentum contributions to the Hamiltonian
constraint:
\begin{equation}
 \jig{11}= \frac{B_1}{G}= -\frac{\sqrt{|\beta|E^x} b_1}{G}\quad,\quad
 \jig{20}= \frac{A_1}{2G}=0\quad,\quad \jig{02}= \frac{B_2}{2G}=
 -b_1b_2\sqrt{|\beta|} \frac{E^{\varphi}}{2G\sqrt{E^x}}\,.
\end{equation}
With these solutions, we obtain $\partial \jig{00}/\partial\Gamma_{\varphi}'=
-{\rm sgn}(\beta)\sqrt{|\beta|E^x}/Gb_1$ from (\ref{B1B2}) and $\partial
\jig{00}/\partial \Gamma_{\varphi}= {\rm sgn}(\beta)\sqrt{|\beta|}(b_2-4 ({\rm
  d}b_1/{\rm d}E^x) (E^x/b_1)) E^{\varphi}\Gamma_{\varphi}/(Gb_1\sqrt{E^x})$
from (\ref{A2B1}), or integrated,
\begin{equation} \label{H00}
 \jig{00}=-\frac{{\rm sgn}(\beta)\sqrt{|\beta|}}{G}\left(\frac{\sqrt{E^x}}{b_1}
  \Gamma_{\varphi}'-\frac{1}{b_1}
\left(\frac{b_2}{2}-\frac{2E^x}{b_1} \frac{{\rm d}b_1}{{\rm d}E^x}\right)
\frac{ E^{\varphi}}{\sqrt{E^x}} \Gamma_{\varphi}^2\right)+ f(E^x)E^{\varphi}\,.
\end{equation}
Comparing with the general form (\ref{myEffectiveHamiltonian}), we
read off
\begin{eqnarray}
 \bar{\alpha}= \sqrt{|\beta|}b_1\quad&,&\quad
 \alpha= \sqrt{|\beta|}b_1b_2\,,\\
\bar{\alpha}_{\Gamma}= {\rm sgn}(\beta)\frac{\sqrt{|\beta|}}{b_1} \quad&,&\quad
 \alpha_{\Gamma}= {\rm sgn}(\beta)\frac{\sqrt{|\beta|}}{b_1}\left(b_2- 
4
 \frac{{\rm d}\log b_1}{{\rm d}\log E^x}\right)\,.
\end{eqnarray}
With these relationships, the correction functions can easily be seen
to satisfy the condition (\ref{alphaGamma}) as well as
$\beta=\bar{\alpha}\bar{\alpha}_{\Gamma}$.

Modifications to the spherically symmetric dynamics are not entirely
determined by the constraint algebra, consistent with the results of
\cite{LTB,LTBII}.  The function $b_1$ is related to the ratio of
$\bar{\alpha}$ to $\bar{\alpha}_{\Gamma}$, and $b_2$ determines how $\alpha$
differs from $\bar{\alpha}$. The $E^x$-dependence of $\jig{00}$ in (\ref{H00})
(which may include a cosmological-constant term) is not fully determined
because $E^x$ is a scalar with no density weight and can, for the purpose of
the constraint algebra, be inserted rather freely in the constraints. In this
feature we can see why the full dynamics is more unique than the spherically
symmetric one: Without symmetry, there is less freedom in the choice of
spatial tensors with the correct transformation properties. Indeed, as we will
see later, spatial transformation properties play an important role for the
regaining procedure. Without spherical symmetry $\Gamma_{\varphi}'$ and
$\Gamma_{\varphi}^2$ would be part of the same contribution $^{(3)}R$, which
cannot be split apart by different correction functions if the spatial
structure of geometry remains unmodified. The case of $\alpha=\bar{\alpha}$
($b_2=1$) and $\alpha_{\Gamma}=\bar{\alpha}_{\Gamma}$ ($b_1$ constant and
therefore $b_1=1$ for it to approach one at large fluxes) is then preferred,
with all corrections determined by the algebraic deformation $\beta$.

\subsection{Legendre transform}

Instead of having to assume $\delta H/\delta\pi^{ij}$ (or $\delta
H/\delta K_x$ and $\delta H/\delta K_{\varphi}$ in spherical symmetry
with triad variables) to be linear in the momenta, it is more general
to treat $\delta H/\delta\pi^{ij}(x)=:v_{ij}(x)$ as a new independent
variable in place of $\pi^{ij}$, and then expand by this newly defined
$v_{ij}$. This change amounts to a Legendre transformation from
$(g_{ij},\pi^{ij})$ with Hamiltonian $H$ to $(g_{ij},v_{ij})$ with
Lagrangian $L=\pi^{ij}v_{ij}-H$, as proposed in
\cite{LagrangianRegained}. We then have the equations
\begin{align}
H &= \pi^{ij} v_{ij}-L & \fund{H}{\met{x'}}\bigg\rvert_{\mom{x}} &=
-\fund{L}{\met{x'}}\bigg\rvert_{\ex{x}}\,. 
\end{align}

There are now two differences to \cite{LagrangianRegained}. First, our
$v_{ij}$ here need not be geometrical extrinsic curvature because of
modifications to space-time geometry. We simply define a new independent
variable $v_{ij}= (\delta N)^{-1}\{g_{ij},H[\delta N]\}$, which we interpret
as the rate of change of the metric, eventually providing time derivatives in
an effective action.  Secondly, we cannot always assume that the Hamiltonian
is local and free of derivatives of $\pi^{ij}$, which would imply that partial
derivatives could be used to compute $v_{ij}$.

Using $v_{ij}$, we write the Poisson bracket of two smeared
Hamiltonian constraints as
\begin{eqnarray}
 \{H[N],H[M]\} &=& \int{\rm d}^3x \frac{\delta H[N]}{\delta g_{ij}(x)}
 v_{ij}(x) M(x)- (N\leftrightarrow M)\\
&=& -\int{\rm d}^3x\int{\rm d}^3y \frac{\delta L(y)}{\delta g_{ij}(x)}
 v_{ij}(x) N(y)M(x)- (N\leftrightarrow M)\\
&=& \int{\rm d}^3x \beta D^i(x) (NM_{|i}-MN_{|i})
\end{eqnarray}
with the local diffeomorphism constraint $D^i$.
%
%
Taking functional derivatives by $N$ and $M$, we arrive at the functional
equation
\begin{equation} \label{FunctionalL}
 \fund{L(x)}{\met{x'}}\ex{x'} + \beta(x)D^{i}(x)\delta_{|i}(x,x')-\vv = 0
\end{equation}
for $L(x)$, which can be solved once an expression for the diffeomorphism
constraint $D^i$ is inserted. With $D^i$ linear in the momenta, a fact which
remains true in the cases of deformed constraint algebras considered here, and
momenta related to functional derivatives of $L$ by $v_{ij}$, a linear
equation for $L$ is obtained. The importance of this consequence of the
Legendre transform has been stressed in \cite{LagrangianRegained}.

If gravity and matter split into independent constrained systems, as realized
for matter constraints independent of the gravitational momentum and in the
absence of derivatives of $\mom{x}$ in $H_{\rm grav}$, equation
(\ref{FunctionalL}) can be derived in an analogous form for the matter part,
just using canonical matter variables and the matter diffeomorphism
constraint. Because the following calculations, integrating the functional
differential equation, are easier for scalar matter, we will first consider
this case as an illustration of the general procedure. As we will see, the
Lagrangian viewpoint provides a new interpretation of conditions of anomaly
freedom found earlier for inverse-triad corrections of a scalar field.

\subsection{Scalar matter}
\label{s:Scalar}

With the classical spatial structure, the Lagrangian density of a scalar field
$\phi$ must be of the form ${\cal L}= \sqrt{\det g} L(\phi,V,\psi)$ where
$V=(\delta N)^{-1}\{\phi,H[\delta N]\}$ is the normal scalar velocity
introduced before and $\psi= g^{ij} \phi_{|i}\phi_{|j}$ is the only remaining
scalar that can be formed from $\phi$ and its derivatives up to a total
derivative order of at most two. Higher derivatives do not appear classically
for equations of motion of second order, but they can easily be introduced by
quantum effects. Higher spatial derivatives, in particular, are a natural
consequence of discretization in loop quantum gravity, which in effective form
combined with a derivative expansion will give rise to derivative terms of
arbitrary orders. Higher time derivatives, on the other hand, follow from
quantum back-reaction. The following considerations for matter assume the
absence of higher-order derivatives, as realized for instance for
inverse-triad corrections and some forms of holonomy corrections.

With the canonical variables of a scalar field and its diffeomorphism
constraint $D^i=p_{\varphi}\phi^{|i}$, equation (\ref{FunctionalL}), adapted
to a scalar field, assumes the form
\begin{equation}
 \frac{\delta L(x)}{\delta\phi(x')} V(x')+ \beta \frac{\partial
   L(x)}{\partial V(x)} \phi^{|i}(x) \delta_{|i}(x,x') -\vv=0\,.
\end{equation}
As in \cite{LagrangianRegained}, we write
\[
 \frac{\delta L(x)}{\delta \phi(x')}= \frac{\partial L(x)}{\partial \phi(x)}
 \frac{\delta\phi(x)}{\delta\phi(x')}+ 2\frac{\partial
   L(x)}{\partial\psi(x)} \phi^{|i}(x) \delta)_{|i}(x,x'
\]
and conclude, taking into account the additional factor of $\beta$, that 
\[
 A^i:= \phi^{|i} \left(\beta\frac{\partial L}{\partial V}+
   2V\frac{\partial L}{\partial\psi}\right)
\]
satisfies the equation $A^i(x)\delta_{|i}(x,x')-\vv=0$, shown in
\cite{LagrangianRegained} to imply $A^i=0$. Thus,
\[
 \beta\frac{\partial L}{\partial V}+
   2V\frac{\partial L}{\partial\psi}=0
\]
and $L$ must be of the form $L(\phi,\psi-V^2/\beta)$. 

This is a concrete indication that the deformed hypersurface-deformation
algebra implies a modification of the usual covariance and of the dispersion
relation of fields: The kinetic term of scalar Lagrangians does not depend on
$\psi-V^2= \stmet^{\mu\nu}\phi_{|\mu}\phi_{|\nu}$ in space-time terms, but has
its time derivatives in $\psi-V^2/\beta$ rescaled by the correction function
$\beta$. Nevertheless, the system is covariant and consistent, albeit with a
deformed notion of covariance as per the constraint algebra (\ref{HH}).  The
dependence of the Lagrangian on the potential remains unrestricted, leaving
the form of some counterterms as introduced in \cite{ConstraintAlgebra} more
open.

It is illustrative to compare this form of the kinetic term with the
one obtained for the matter Hamiltonian in a consistent deformation
\cite{ConstraintAlgebra}. One begins with a matter Hamiltonian density
of the form
\[
 H= \nu \frac{p_{\phi}^2}{2\sqrt{\det g}}+ \frac{1}{2}\sigma \sqrt{\det
   g}\psi+ \sqrt{\det g} W(\phi)
\]
with metric factors corrected by inverse-triad corrections $\nu$ and $\sigma$,
and some potential $W(\phi)$. The corresponding Lagrangian density, with
$V=\nu p_{\phi}/\sqrt{\det g}$, takes the form
\[
 L= \sqrt{\det g}\left(\frac{V^2}{2\nu}-\frac{\sigma\psi}{2}-
   W(\phi)\right)= -\sqrt{\det g}\frac{\sigma}{2}\left(\psi-
\frac{V^2}{\beta}\right)- \sqrt{\det g} W(\phi)
\]
with the kinetic dependence as derived above, provided that
$\beta=\nu\sigma$. This condition, as derived in \cite{ConstraintAlgebra} for
linear inhomogeneities around isotropic models, is exactly one of the
requirements for anomaly freedom to ensure a closed constraint algebra of the
form (\ref{HH}) for inverse-triad corrections with $\beta=\bar{\alpha}^2$ from
the gravitational constraint. The Lagrangian view clearly shows how this
condition of anomaly cancellation is necessary to ensure a (deformed)
covariant kinetic term in the action. With the same corrections in
d'Alembertians, propagation speeds of massless matter and gravitational waves
naturally agree, as explicitly shown for electromagentic waves in
\cite{tensor}.

From the new derivation of corrected scalar Lagrangians in this paper, we must
expect corrections in matter terms also if $\beta$ results from holonomy
corrections, provided they can be consistently implemented. Explicit examples
for holonomy modifications required in matter terms have already been found in
\cite{ScalarHolEv,ScalarHol}. However, in a scalar Hamiltonian quantized by
the methods of loop quantum gravity \cite{QSDV} we do not expect holonomy
corrections. Consistent formulations of holonomy corrections in the presence
of matter therefore seem to encounter stronger difficulties than inverse-triad
corrections. Another peculiar feature can be seen by recalling that $\beta$
for holonomy corrections can turn negative. The modified d'Alembertian
$\psi-V^2/\beta$ then becomes one of Euclidean signature, or a 4-dimensional
Laplacian, and fields no longer propagate. Also this property can explicitly
be seen in the wave equations of \cite{ScalarHol} (but not in
\cite{ScalarHolEv} where a gauge-fixing has veiled this effect). We will
discuss further consequences of this new form of signature change in
Sec.~\ref{s:Sig}.

\subsection{Gravitational part}

As in the case of scalar matter, we begin our discussion of the gravitational
part by inserting the explicit expression of the diffeomorphism constraint in
the general equation (\ref{FunctionalL}): In particular,
\begin{equation} \label{Hi}
\beta(x)D^{i}(x)\delta_{|i}(x,x')-\vv =
-2\beta(x)\momd{|j}{x}\delta_{|i}(x,x')-\vv \,.
\end{equation}
We then proceed as in the example of spherical symmetry:
We multiply this expression by two test functions $a(x)$ and $b(x')$ and
integrate over $x$ and $x'$,
observing that some terms symmetric in $a$ and $b$ cancel.
After several steps, integrating by parts, discarding total derivatives and
using the symmetry of $\pi^{ij}$, we arrive at
\begin{equation}
\int \dif x \left[ 2 \mom{x}
  \beta_{|j}(x)\left(a(x)b_{|i}(x)-a_{|i}(x)b(x)\right) + 2 \mom{x}
  \beta(x)\left(a(x)b_{|ij}(x)-a_{|ij}(x)b(x)\right)\right]
\end{equation}
from the right-hand side of (\ref{Hi}).
Functional derivatives with respect to $a(y)$ and $b(z)$ give
\begin{align}
&\int \dif x \left[ 2 \mom{x} \beta_{|j}(x) \left(\delta(x,y)\delta_{|i}(x,z)-\delta_{|i}(x,y)\delta(x,z)\right)\right.\notag \\
&+\left. 2 \mom{x} \beta(x) \left(\delta(x,y)\delta_{|ij}(x,z)-\delta_{|ij}(x,y)\delta(x,z)\right)\right]\notag \\
&= 2 \mom{y} \beta_{|j}(y) \delta_{|i}(y,z) + 2 \mom{y} \beta(y) \delta_{|ij}(y,z)- (y \leftrightarrow z)\notag\\
&= 2 \pard{L(y)}{\ex{y}} \beta_{|j}(y) \delta_{|i}(y,z) + 2 \pard{L(y)}{\ex{y}}\beta(y)\delta_{|ij}(y,z) - (y \leftrightarrow z)
\end{align}
if no spatial derivatives of $v_{ij}$ appear in the corrections and the
Lagrangian, such that $\pi^{ij}(y)= \delta L/\delta v_{ij}(y)= \partial
L(y)/\partial v_{ij}(y)$.  In combination with (\ref{FunctionalL}), we have
\begin{equation} \label{modLagrangianConstraint}
\fund{L(x)}{\met{x'}}\ex{x'} + 2 \beta_{|j}(x)\pard{L(x)}{\ex{x}}\delta_{|i}(x,x') + 2 \beta(x)\pard{L(x)}{\ex{x}}\delta_{|ij}(x,x') - \vv = 0\,.
\end{equation}
In cases of derivative expansions of non-local terms in $v_{ij}$, we use
\begin{equation} \label{modLagrangianNonLocal}
 \frac{\delta L(x)}{\delta g_{ij}(x')} v_{ij}(x')\delta(x,x')+
 2\beta_{|j}(x)\frac{\delta L(x)}{\delta v_{ij}(x')}
 \delta_{|i}(x,x')+2\beta(x) \frac{\delta L(x)}{\delta v_{ij}(x')}
 \delta_{|ij}(x,x')- (x\leftrightarrow x')=0
\end{equation}
and write 
\begin{equation}
 \frac{\delta L(x)}{\delta v_{ij}(x')}= \frac{\partial L(x)}{\partial
   v_{ij}(x')} \delta(x,x')+ \frac{\partial L(x)}{\partial v_{ij|k}(x')}
 \delta_{|k}(x,x')+\cdots
\end{equation}

\subsubsection{Expansion}

In spherical symmetry, it turned out to be useful to consider expansion
coefficients by the momenta $K_x$ and $K_{\varphi}$.
As the next crucial step in solving the functional equation, we expand both
$L$ and $\beta$ as series in powers of the normal change of the
metric, $v_{ij}$:
\begin{align} 
L(x) =& \sum_{n = 0}^{\infty} L^{i_{1}j_{1}\ldots
  i_{n}j_{n}}[g^{kl}]\ex[i_{1}j_{1}]{x}\ldots \ex[i_{n}j_{n}]{x} \label{Lexp}\\
\beta(x) =&  \sum_{n = 0}^{\infty} \beta^{i_{1}j_{1}\ldots
  i_{n}j_{n}}[g^{kl}]\ex[i_{1}j_{1}]{x}\ldots \ex[i_{n}j_{n}]{x} \label{betaexp}
\end{align}
assuming for now local functions without spatial derivatives. (See
Sec.~\ref{s:NonLocal} for non-locality.)  The expansion of $\beta$ allows us
to deal with inverse-triad corrections and local holonomy corrections at the
same time. Holonomy corrections will then not appear as periodic functions
such as $\sin(\delta K_{\varphi})/\delta$ for $K_{\varphi}$ in spherical
symmetry, but as perturbative terms of a power series in $K_{\varphi}$. Such
an expansion is more consistent with the perturbative nature of these
higher-order corrections, which are expected in a similar form from
higher-curvature terms or quantum back-reaction. Including all terms in a
power series of $\sin(\delta K_{\phi}/\delta)$, even tiny ones at high orders,
but ignoring quantum back-reaction would not be consistent. An expansion also
makes it more clear how terms of higher order in $v_{ij}$ can be combined with
higher spatial derivatives of the metric.

We insert these expansions into (\ref{modLagrangianConstraint}) and first set
$\ex{x}=0$ to obtain  
\begin{equation}
  2 \Gb \beta^{\emptyset}_{|j}(x)\delta_{|i}(x,x') + 2 \Gb
\beta^{\emptyset}(x) \delta_{|ij}(x,x') - \vv = 0 \,.
\end{equation}
We multiply by test functions $a(x)$ and $b(x')$ and integrate over
$x$ and $x'$,
drop total divergences and terms that vanish due to the symmetry of
indices of \Gb, cancel some other terms and are left with 
\begin{equation}
  \int \dif x \Gbd{|j}\beta^{\emptyset}(a_{|i} b-a b_{|i}) = 0\,.  
\end{equation}
Since $a$, $b$, $a_{|i}$ and $b_{|i}$ can be chosen independently, we conclude
that $\Gbd{|j}\beta^{\emptyset} = 0$.  Note that $\ao \neq 0$ generically, so
that we have
\begin{equation}\label{FirstOrderGCondition}
\Gbd{|j} = 0\,.
\end{equation}

We return to equation \eqref{modLagrangianConstraint}, do a functional
differentiation with respect to \ex[{kl}]{z} and then set \ex{x} to zero
everywhere. With the notation
\begin{equation}
\delta^{kl}_{ab}(x,z)=\frac{1}{2}(\delta^{k}_{a} \delta^{l}_{b} + \delta^{l}_{a} \delta^{k}_{b})\mspace{1mu}\delta\mspace{-1mu}(x,z)
\end{equation}
we have
\begin{align}
\fund{L^{\emptyset}(x)}{\met[kl]{x'}}&\delta(x',z) + 4L^{ijab}(x)\aod{|j}\delta_{|i}(x,x')\delta^{kl}_{ab}(x,z)\notag\\
 &+ 2\Gb\delta_{|i}(x,x')\left(\aad{ab}{|j}\delta^{kl}_{ab}(x,z)+\aan{ab}\delta^{kl}_{ab|j}(x,z)\right)\notag\\
 &+ 4L^{ijab}\delta^{kl}_{ab}(x,z)\delta_{|ij}(x,x')+ 2\Gb\aan{ab}\delta^{kl}_{ab}\delta_{|ij}(x,x') - \vv \\
 = -\fund{L^{\emptyset}(x')}{\met[kl]{x}}&\delta(x,z) + \left(4L^{ijkl}(x)\aod{|j} + 2\Gb\aad{kl}{|j}\right)\delta_{|i}(x,x')\delta(x,z)\notag\\
 &+ 2\Gb\aan{kl}\delta_{|j}(x,z)\delta_{|i}(x,x')\notag\\
 &+ \left(2\Gb\aan{kl} + 4L^{ijkl}(x)\ao\right)\delta_{|ij}(x,x')\delta(x,z) - \vv =0. \label{1Klvlexp}
\end{align}
We use
\begin{align} \label{G2A1exp}
2\Gb\aan{kl}\delta_{|j}(x,z)&\delta_{|i}(x,x') = \left(2\Gb\aan{kl}\delta(x,z)\delta_{|i}(x,x')\right)_{|j}\notag \\
&{}- 2\Gbd{|j}\aan{kl}\delta(x,z)\delta_{|i}(x,x') - 2\Gb\aad{kl}{|j}\delta(x,z)\delta_{|i}(x,x')\notag\\
&{}-2\Gb\aan{kl}\delta(x,z)\delta_{|ij}(x,x') \,,
\end{align}
drop the total divergence term in \eqref{G2A1exp}, and insert $\Gbd{|j}=0$
from (\ref{FirstOrderGCondition}):
\begin{equation} \label{1Ksub}
\left(-\fund{L^{\emptyset}(x')}{\met[kl]{x}} + 4L^{ijkl}(x)\aod{|j}\delta_{|i}(x,x') + 4L^{ijkl}\ao\delta_{|ij}(x,x')\right)\delta(x,z) - \vv = 0.
\end{equation}
This equation can be solved as in \cite{LagrangianRegained} where $\beta^{\emptyset}=1$: define
\begin{equation}
A^{ij}(x,x') = \fund{L^{\emptyset}(x)}{\met{x'}} - 
4L^{ijkl}(x')\left(\aod{|l}(x')\delta_{|k}(x',x) + \ao(x')\delta_{|kl}(x',x)\right)
\end{equation}
and rewrite \eqref{1Ksub} as
\begin{equation}
A^{ij}(x,x')\delta(x',z) - A^{ij}(x',x)\delta(x,z) = 0\,.
\end{equation}
Integrating over $x'$, we find
\[
A^{ij}(x,x'') = F^{ij}(x)\delta(x,x'') \quad\mbox{with}\quad
F^{ij}(x) = \int{\dif^{3} x' A^{ij}(x',x) }
\]
a function of only one variable,
and thus
\begin{equation} \label{G0}
\fund{L^{\emptyset}(x)}{\met{x'}} = F^{ij}(x)\delta(x,x') + 4L^{ijkl}(x')\left(\aod{|l}(x')\delta_{|k}(x',x) + \ao(x')\delta_{|kl}(x',x)\right)\,.
\end{equation}

\subsubsection{Coefficients}

As a spatial scalar density, $L^{\emptyset}$ can depend on the metric
and its spatial derivatives only via the metric itself and suitable
contractions of products of the spatial Riemann tensor. To second order in
spatial derivatives,
\begin{equation}
L^{\emptyset}(x) = L^{\emptyset}(\met{x},\,{}^{(3)}\!R_{ij}(x))\,,
\end{equation}
a fact, used in \cite{LagrangianRegained}, that remains true in the deformed
case with our assumption that the spatial part of the algebra stays
classical.  Define
\begin{equation} \label{phiPhi}
\varphi^{ij} = \pard{L^{\emptyset}(g_{kl},\,{}^{(3)}\!R_{kl})}{g_{ij}} \qquad \Phi^{ij} = \pard{L^{\emptyset}(g_{kl},\,{}^{(3)}\!R_{kl})}{\,{}^{(3)}\!R_{ij}}
\end{equation}
and write
\begin{align}
\delta L^{\emptyset} =& \left(\varphi^{ij} + \frac{1}{2}\,{}^{(3)}\!R^{i}{}_{kl}\,{}^{j}\ph{kl} + \frac{1}{4}\,{}^{(3)}\!R^{i}_{k}\ph{kj} + \,{}^{(3)}\!R^{j}_{k}\ph{ki} \right)\delta\metd{ij}\notag\\
&+{}\frac{1}{4}\left(\ph{ik}\um{jl} + \ph{il}\um{jk} + \ph{jk}\um{il} + \ph{jl}\um{ik} - 2\ph{ij}\um{kl} -2\ph{kl}\um{ij}\right)\delta\metd{ij|kl}\,.
\end{align}
From (\ref{G0}), we also have
\begin{align}
\delta L^{\emptyset} =& \delta\metd{ij}\left(F^{ij} + 4L^{ijkl}_{|lk}\ao + 4L^{ijkl}_{|l}\aod{|k}\right)\notag\\
&+{} \delta\metd{ij|k}\left(8L^{ijkl}_{|l}\ao + 4L^{ijkl}\aod{|l}\right)
+{} \delta\metd{ij|kl}\left(4L^{ijkl}\ao\right)\,.
\end{align}
Comparing the various coefficients, we get
\begin{align}
&L^{ijkl}\ao = \frac{1}{16}\left(\ph{ik}\um{jl} + \ph{il}\um{jk} + \ph{jk}\um{il} + \ph{jl}\um{ik} - 2\ph{ij}\um{kl} -2\ph{kl}\um{ij}\right)\label{1KexpG4P}\\
&2L^{ijkl}_{|l}\ao + L^{ijkl}\aod{|l} = 0\label{G2Condition}\\
&F^{ij} + 4L^{ijkl}_{|lk}\ao + 4L^{ijkl}_{|l}\aod{|k} = \varphi^{ij} +
\frac{1}{2}\,{}^{(3)}\!R^{i}{}_{kl}\,{}^{j}\ph{kl} +
\frac{1}{4}\,{}^{(3)}\!R^{i}_{k}\ph{kj} +
\frac{1}{4}\,{}^{(3)}\!R^{j}_{k}\ph{ki}\,. \label{Fphi}
\end{align}
Thus,
\begin{equation}
  0=2L^{ijkl}_{|l}\ao + L^{ijkl}\aod{|l} = -\beta^{\emptyset}_{|l}
  L^{ijkl}+ 2(L^{ijkl}\beta^{\emptyset})_{|l}\,.
\end{equation}
We compute each term using \eqref{1KexpG4P}, and write
\begin{align}\label{PhiMetAlphZero}
0 ={}& -\frac{\aod{|l}}{16\ao}\left(\ph{ik}\um{jl} + \ph{il}\um{jk} + \ph{jk}\um{il} + \ph{jl}\um{ik} - 2\ph{ij}\um{kl} - 2\ph{kl}\um{ij}\right)\notag\\
&+{} \frac{1}{8}\left(\Phi^{ik|j} + \um{jk}\phdl{il} + \Phi^{jk|i} + \um{ik}\phdl{jl} - 2\Phi^{ij|k} - 2\phdl{kl}\um{ij}\right)\,.
\end{align}
We contract this with $\metd{ij}$,
use $\kd{i}{i} = 3$, and denote $\Phi^{i}_{i}$ as $\Phi$:
\begin{equation}
\frac{\aod{|l}}{8\ao}\left(\ph{kl} + \Phi\um{kl}\right) - 
\frac{1}{4}\left(\phdl{kl} + \Phi^{ij|k}\metd{ij}\right) = 0\,.
\end{equation}
Note that $\Phi^{ij|k}\metd{ij} = \Phi^{|k} = (\Phi\um{kl})_{|l}$.
With $\ph{kl} + \Phi\um{kl}$ denoted as $\bar{\Phi}^{kl}$,
\begin{equation} \label{Phibardiff}
0=\frac{\aod{|l}}{8\ao}\bph{kl} - \frac{1}{4}\bphdl{kl}
=\frac{1}{4}\sqrt{|\ao|}\left(\frac{\aod{|l}{\rm
    sgn}(\ao)}{2|\ao|^{\frac{3}{2}}}\bph{kl} -
|\ao|^{-\frac{1}{2}}\bphdl{kl} \right)=
-\frac{1}{4}\sqrt{|\ao|}
 \left(|\ao|^{-\frac{1}{2}}\bph{kl}\right)_{|l} \,.
\end{equation}

Again maintaining our assumption of an unmodified spatial structure, the only
covariantly constant 2-tensors constructed from the metric and its derivatives
up to second order are the metric itself and the spatial Einstein
tensor. Noting the density weight one of $\bar{\Phi}^{kl}$, inherited from
$L^{\emptyset}$, we conclude that
\begin{equation}
\frac{\bph{kl}}{\sqrt{|\ao|}} = A\sqrt{\det g}\left(\,{}^{(3)}\!R^{kl} - 
\frac{1}{2}\,{}^{(3)}\!R\um{kl}\right) + B\sqrt{\det g}\um{kl}
\end{equation}
where $A$ and $B$ are constants.
This gives
\begin{equation}
\ph{kl} = A\sqrt{|\ao|\det g}\left(\,{}^{(3)}\!R^{kl} - 
\frac{3}{8}\,{}^{(3)}\!R\um{kl}\right) + \frac{B}{4}\sqrt{|\ao|\det g}\um{kl}\,.
\end{equation}
Inserting this into \eqref{PhiMetAlphZero}, we find, after cancelling terms,
that 
\begin{multline}
\frac{A\sqrt{|\ao|\det g}}{8}\bigg[\,{}^{(3)}\!R^{ik}\um{jl} +
\,{}^{(3)}\!R^{il}\um{jk}
 + \,{}^{(3)}\!R^{jk}\um{li} + \,{}^{(3)}\!R^{jl}\um{ik} -
 2\,{}^{(3)}\!R^{ij}\um{kl} 
- 2\,{}^{(3)}\!R^{kl}\um{ij}\\
-{}\frac{3}{8}\,{}^{(3)}\!R\left(2\um{ik}\um{jl} + 2\um{jk}\um{il} - 
4\um{ij}\um{kl}\right)\bigg]_{|l} = 0\,.
\end{multline}
For this to be satisfied for general metrics, we must set
$A=0$. Writing $B =\frac{1}{4\pi G}$,
\begin{equation}
\ph{kl} = \frac{1}{16\pi G}\sqrt{|\ao|\det g}\um{kl}\,. 
\end{equation}
Then, from (\ref{1KexpG4P})
\begin{align}
L^{ijkl} &{}= \frac{1}{16^2\pi G\ao}\bigg(\sqrt{|\ao|\det g}\um{ik}\um{jl} + 
\sqrt{|\ao|\det g}\um{il}\um{jk} + \sqrt{|\ao|\det g}\um{jk}\um{il}\notag\\
&\mspace{80mu}+{} \sqrt{|\ao|\det g}\um{jl}\um{ik} - 
2\sqrt{|\ao|\det g}\um{ij}\um{kl} - 2\sqrt{|\ao|\det
  g}\um{kl}\um{ij}\bigg)\notag\\ 
&{}= \frac{\sqrt{\det g}{\rm sgn}\ao}{64\pi
  G\sqrt{|\ao|}}\bigg(\um{i(k}\um{l)j}  - \um{ij}\um{kl}\bigg)\,. \label{Lijkl}
\end{align}
We also have
\begin{equation}
\pard{L^{\emptyset}(\metd{kl},\,{}^{(3)}\!R_{kl})}{\,{}^{(3)}\!R_{ij}} = 
\ph{ij} = \frac{1}{16\pi G}\sqrt{|\ao|\det g}\um{ij}
\end{equation}
from the definition (\ref{phiPhi}).
We integrate this to get
\begin{equation} \label{L0}
L^{\emptyset} = \frac{1}{16\pi G}\sqrt{\det g}\left(\sqrt{|\ao|}\,{}^{(3)}\!R +
f(g)\right)
\end{equation}
where, for a scalar density, $f(g)= -2\lambda$ must be a constant, the
cosmological constant. (The previous equations do not determine $f(g)$ because
it would follow from $\varphi^{ij}$ according to (\ref{phiPhi}), which by
(\ref{Fphi}) is related to the free function $F^{ij}$.)

Combining (\ref{Lijkl}) and (\ref{L0}), the regained Lagrangian up to
second order is
\begin{equation} \label{EffAc}
 L=\frac{\sqrt{\det g}}{16\pi G} \left(\frac{{\rm sgn} \ao}{\sqrt{|\ao|}} 
 \frac{v_{ij}v^{ij}- v^i_i v^j_j}{4}
 +  \sqrt{|\ao|} \,{}^{(3)}\!R-2\lambda\right) \,.
\end{equation}
For $\ao=1$, the classical Lagrangian is recovered with $v_{ij}=2K_{ij}$
related to extrinsic curvature. But already to second order in derivatives,
loop quantum gravity implies corrections to the Lagrangian from inverse-triad
corrections with $\ao\not=1$, a property that cannot be mimicked by any form
of higher-curvature effective actions. Also holonomy corrections cannot
provide a similar modification because they always come with higher powers of
$v_{ij}$.  Inverse-triad corrections can thus easily be distinguished from
other quantum effects. (Holonomy corrections can provide similar modifications
if the $v_{ij}$ expansion is resummed; see Sec.~\ref{s:HighDens}.)

The correction function $\beta^{\emptyset}[g_{ij}]$ relevant for these
corrections must be scalar, which is not possible classically if only the
metric can be used. For this reason, the full dynamics is more unique than the
spherically symmetric one, where $E^x$ is a scalar metric component without a
density weight in the reduced model. In an effective formulation of quantum
gravity, additional quantities become available that explicitly refer to
properties of an underlying state, such as the discreteness scale in loop
quantum gravity. It is then possible to construct non-trivial scalars of
density weight zero by referring to the metric and state parameters, such as
elementary fluxes \cite{ConstraintAlgebra}.

Compared with the results in spherical symmetry, the full effective action is
more unique, as already discussed. Other properties of the corrections are,
however, very similar: The correction function $\beta$ features in the same
way in the curvature potential. Also the kinetic term is corrected in the same
way, if we only note that a factor of $\sqrt{|\ao|}$ was obtained in spherical
symmetry, where we used momenta $K_x$ and $K_{\varphi}$ instead of the normal
change $v_{ij}$ of the metric. If we substitute the normal changes $\delta
H/\delta K_x$ and $\delta H/\delta K_{\varphi}$ for $K_x$ and $K_{\varphi}$ in
spherical symmetry, we also obtain a kinetic term divided by
$\sqrt{|\ao|}$. The sign of $\ao$ appears in different places in our
expressions for spherical symmetry and the full theory, but the relative sign
between the curvature and the kinetic terms is the same. The absolute
placement of the sign is ambiguous because in the derivations it first appears
in derivatives, for instance when we introduce $B_1$ after (\ref{H00diff}), or
in (\ref{Phibardiff}).

\subsubsection{Higher orders}
\label{s:High}

To second order, $\Phi^{kl}$ determines both $L^{ijkl}$ from (\ref{1KexpG4P})
and $\partial L^{\emptyset}/\partial \,{}^{(3)}\!R_{ij}$ from (\ref{phiPhi}),
ensuring that time derivatives of $g_{ij}$ and spatial Ricci contributions are
combined to space-time covariant curvature terms. The same interplay is
repeated for higher orders in the $v$-expansion, although with an increasing
number of terms.

For the next order, as an example, we start again from
\eqref{modLagrangianConstraint} and gather all terms which are
quadratic in \ex{x} and its derivatives.
\begin{align}
\fund{L^{ab}(x)}{\met{x'}}&\ex[ab]{x}\ex{x'} + 6L^{abcdij}\ex[ab]{x}\ex[cd]{x}\aod{|j}\delta_{|i}(x,x')\notag\\
+{}&{} 4L^{abij}(x)\ex[ab]{x}\left(\aan{ef}\ex[ef]{x}\right)_{|j}\delta_{|i}(x,x') + 2\Gb\left(\abn{cdef}\ex[cd]{x}\ex[ef]{x}\right)_{|j}\delta_{|i}(x,x')\notag\\
+{}&{} 6L^{abcdij}\ex[ab]{x}\ex[cd]{x}\ao\delta_{|ij}(x,x') + 4L^{abij}\ex[ab]{x}\aan{cd}\ex[cd]{x}\delta_{|ij}(x,x')\notag\\
+{}&{} 2\Gb\abn{cdef}\ex[cd]{x}\ex[ef]{x}\delta_{|ij}(x,x') - \vv \,.
\end{align}
We multiply this by two test functions, $a(x)$ and $b(x')$ and integrate over
$x$ and $x'$. After integrating by parts, discarding total divergences,
removing terms that disappear due to the symmetry and anti-symmetry of various
indices, and using \eqref{FirstOrderGCondition}, we arrive at
\begin{align}
\iint \dif x \dif x' &{}\left(\fund{L^{ab}(x)}{\met{x'}}-\fund{L^{ij}(x')}{\met[ab]{x}}\right)\ex[ab]{x}\ex{x'}a(x)b(x')\notag\\
-{}&{}\int \dif x \left(6L^{abcdij}\ex[ab]{x}\ex[cd]{x}\right)_{|j}\ao\left(a(x)b_{|i}(x)-a_{|i}(x)b(x)\right)\notag\\
-{}&{}\int \dif x \left(4L^{abij}\ex[ab]{x}\right)_{|j}\aan{cd}\ex[cd]{x}\left(a(x)b_{|i}(x)-a_{|i}(x)b(x)\right) = 0.
\end{align}
Since $v_{ab}(x)$, $v_{ab|j}(x)$, $a(x)$, $b(x)$, $a_{|i}(x)$ and $b_{|i}(x)$
can all be varied independently, we arrive at the following three conditions:
First, setting $\left(a(x)b_{|i}(x)-a_{|i}(x)b(x)\right) = 0$, we get
\begin{equation}
\left(\fund{L^{ab}(x)}{\met{x'}}-\fund{L^{ij}(x')}{\met[ab]{x}}\right)\ex[ab]{x}\ex{x'}a(x)b(x') = 0\,.
\end{equation}
Following the arguments in \cite{LagrangianRegained}, we see that this
eventually implies
\begin{equation}
\fund{L^{ab}(x)}{\met{x'}}-\fund{L^{ij}(x')}{\met[ab]{x}} = 0.\label{FuncCurl}
\end{equation}
This equation restricts the form of terms linear in $v_{ij}$ in the action,
which are absent anyway if the theory is time-reversal invariant.
Then setting $v_{ab|j}(x) = 0$,
\begin{equation}
6L^{abcdij}_{|j}\ao + 4L^{ijab}_{|j}\aan{cd} = 0\,.\label{G3Cond1}
\end{equation}
And finally:
\begin{equation}
12L^{abcdij}\ao + 4L^{abij}\aan{cd} = 0.\label{G3Cond2}
\end{equation}

We relabel indices, multiply \eqref{G3Cond2} with $\aod{|j}$ and use
\eqref{G2Condition} to rewrite it.
\begin{align}
12L^{ijklmn}\ao\aod{|j} + 4L^{ijkl}\aan{mn}\aod{|j}
=12L^{ijklmn}\ao\aod{|j} - 8L^{ijkl}_{|j}\aan{mn}\ao &\notag\\
=12L^{ijklmn}\ao\aod{|j} - 8L^{mnij}_{|j}\aan{kl}\ao &{}= 0 \,.
\end{align}
(We use $L^{ijklmn}= L^{ijmnkl}$, referring to the definition in
(\ref{Lexp}).)  Using \eqref{G3Cond1}, we can write
$24L^{ijklmn}_{|l}(\ao)^{2} + 24L^{ijklmn}\ao\aod{|l} = 0$.  Generically, $\ao
\neq 0$, and so we have $\left(L^{ijklmn}\ao\right)_{|l} = 0$ solved by the
classical covariantly constant quantities with the corresponding index
structure, divided by $\beta^{\emptyset}$.

The third order in $v_{ij}$ will therefore have terms with a factor of $1/\ao$
times corresponding orders possible for higher-curvature actions, while the
quadratic order had a factor of $1/\sqrt{|\ao|}$, and the zeroth order a
factor of $\sqrt{|\ao|}$. The same pattern is repeated at higher orders in the
$v$-expansion: To order $n$ in $v_{ij}$, we have terms as in higher-curvature
actions but multiplied with $|\ao|^{(1-n)/2}$. To see this, we notice that
Eq.~(\ref{modLagrangianConstraint}), when expanded by powers of $v_{ij}$, has
a first term which contains expansion coefficients of $L^{ij\cdots}$ two
orders lower than the rest, which are all multiplied with $\ao$. If we use the
equation to derive the $L$-coefficients by recurrence, we solve for a
coefficient two orders higher by dividing by $\ao$. Starting with zeroth order
in $v_{ij}$ of magnitude $\sqrt{|\ao|}$ in the prefactor, the quoted orders
follow. (If the corrected theory is not time-reversal invariant and odd orders
appear in the $v$-expansion, the same powers of $|\ao|$ per order are
obtained.)

\subsubsection{Non-locality}
\label{s:NonLocal}

So far, we have assumed only a local dependence on $v_{ij}$, with no spatial
derivatives of $v_{ij}$ that would otherwise be implied by a derivative
expansion. In the classical case, locality follows from the relation of
$v_{ij}$ to extrinsic curvature, but it can easily be violated by some of the
correction functions in quantum gravity.

In an effective action, non-locality usually makes itself noticeable in a
derivative expansion of the fields. The basic equation (\ref{FunctionalL}) is
valid also for non-local theories, without explicit terms with spatial
derivatives of $v_{ij}$. However, (\ref{modLagrangianConstraint}) must be
replaced by (\ref{modLagrangianNonLocal}), and the general expansions
(\ref{Lexp}) and (\ref{betaexp}) must also include terms with spatial
derivatives of $v_{ij}$. We now define
\begin{eqnarray} \label{LK}
L(x)&=& \sum_{n=0}^{\infty} \sum_{N_1,\ldots,N_n=0}^{\infty} 
L^{(i_1,j_1,k_1^{(1)},\ldots,k_1^{(N_1)}),\ldots,
  (i_n,j_n,k_n^{(1)},\ldots,k_n^{(N_n)})}[g_{ij}]\\
&&\times v_{i_1j_1|k_1^{(1)}\cdots k_1^{(N_1)}}\cdots 
v_{i_nj_n|k_n^{(1)}\cdots k_n^{(N_n)}}\nonumber\\
\beta(x)&=& \sum_{n=0}^{\infty} \sum_{N_1,\ldots,N_n=0}^{\infty} 
\beta^{(i_1,j_1,k_1^{(1)},\ldots,k_1^{(N_1)}),\ldots,
  (i_n,j_n,k_n^{(1)},\ldots,k_n^{(N_n)})}[g_{ij}]\\
&&\times v_{i_1j_1|k_1^{(1)} \cdots k_1^{(N_1)}}\cdots 
v_{i_nj_n|k_n^{(1)}\cdots k_n^{(N_n)}}\nonumber\,.
\end{eqnarray}
Derivative terms in the expansion of $\beta$ then require new terms in the
Lagrangian that contain spatial derivatives. Going through the recurrence, an
order $n$ in the $v$-expansion again receives a coefficient of
$|\ao|^{(1-n)/2}$.

In this context, we can distinguish between two expansions of the action, one
by powers of $v_{ij}$ and its spatial derivatives as in (\ref{LK}), and one by
the total order of derivatives. The total order of derivatives is the crucial
one for a comparison with higher-curvature terms in an effective action, which
come arranged by the order of time and space derivatives. With $v_{ij}$
related to the normal change of the metric, it counts as a derivative (by
time) of order one. A term of $v_{i_1j_1|k_1^{(1)}\cdots k_1^{(N_1)}}$ counts
as a derivative of order $1+N_1$, and therefore a general expansion term in
(\ref{LK}) with coefficient $L^{(i_1,j_1,k_1^{(1)},\ldots,k_1^{(N_1)}),\ldots,
  (i_n,j_n,k_n^{(1)},\ldots,k_n^{(N_n)})}$ counts as a derivative of order
$\sum_{i=1}^n (1+N_i) =n+\sum_{i=1}^n N_i$.  Terms of the same $v$-order $n$,
that is with the same number of factors of $v_{ij}$ or its spatial
derivatives, have different derivative orders of at least $n$. If we
reorganize the expansion by derivative orders $N$, keeping track of
$\ao$-factors that depend only on the $v$-order, we obtain effective-action
terms of the schematic form
\begin{eqnarray*}
&& |\beta^{\emptyset}|^{(1-N)/2} v^N+
|\beta^{\emptyset}|^{(2-N)/2} (v^{N-1}g'+v^{N-2}v')\\
&&\quad+
|\beta^{\emptyset}|^{(3-N)/2} \left(v^{N-2}(g''+(g')^2)+ v^{N-3}(v''+v'g')+ 
v^{N-4}(v')^2\right)+\cdots\,.
\end{eqnarray*}
The highest power of $1/\sqrt{|\ao|}$ for a given derivative order is always
obtained for the term $v^N$ free of spatial derivatives.  For small $\ao$,
time derivatives in a derivative or curvature expansion are dominant.

\section{Applications and conclusions}

One of the main results of this paper, of general importance for loop quantum
gravity, follows from the effective action (\ref{EffAc}), valid to second
order in extrinsic curvature. Although we did allow for holonomy and
higher-curvature corrections as well, only inverse-triad corrections are
active at this order. This result is an independent confirmation, in addition
to \cite{InflObs,InflConsist,InflTest}, that inverse-triad corrections can be
much more significant than higher-curvature and holonomy corrections, both of
which occur only at higher orders in $v_{ij}$ and are of the tiny size
$\ell_{\rm P}^2/\ell_{\cal H}^2$ throughout most of nearly isotropic cosmology
with the Hubble distance $\ell_{\cal H}$. Our calculations show, for the first
time, how different quantum effects in loop quantum gravity without any
symmetry assumptions can be included all at once, but still show their own
characteristic consequences. The complete correction function $\beta$ in the
constraint algebra may contain contributions from both inverse-triad and
holonomy corrections, including non-local effects, but it is only the
$v$-independent part $\ao$ which appears at second order of the effective
action. This coefficient is affected by inverse-triad corrections, which
therefore present the most important modification of the classical dynamics
unless curvature is extremely large. Holonomy corrections, on the other hand,
modify terms of higher order in $v_{ij}$; they mix with higher-curvature terms
and can rarely be used in isolation. Moreover, U(1) calculations of
inverse-triad correction functions are reliable because non-Abelian features
would change merely the higher-$v$ behavior.

The clear separation of some of the corrections allows us to discuss
their cosmological consequences in very general terms.

\subsection{Enhanced BKL scenario and the absence of singularities in
  consistent loop quantum gravity}

All $v_{ij}$-terms in the effective action (\ref{EffAc}), to all orders, have
at least one additional factor of $1/\ao$ compared with the spatial curvature
term at zeroth order (or a factor of $1/\sqrt{|\ao|}$ if there are linear
terms in $v_{ij}$, breaking time-reversal invariance). At higher orders, as
shown in Secs.~\ref{s:High} and \ref{s:NonLocal}, $v_{ij}$-terms free of
spatial derivatives have at least an additional factor of $1/\sqrt{|\ao|}$
compared to spatial-derivative terms of the same derivative order. When $\ao$
is very small, all spatial derivatives and curvature potentials are suppressed
compared with the normal change of the metric in $v_{ij}=N^{-1}
\{g_{ij},H[N]\}$. Inverse-triad corrections, computed in Abelian models
\cite{DegFull}, imply that $\ao$ approaches zero for vanishing components of
the densitized triad, right at classical singularities. As we approach such a
singularity, quantum corrections become stronger, which could altogether stop
the evolution down to smaller volumes. If this is the case, the singularity is
resolved. However, such ``bounces'' have been difficult to generalize beyond
the simple models in which they can be realized explicitly (see also
Sec.~\ref{s:Sig} below), and therefore it is not guaranteed that vanishing
components of the densitized triad can always be avoided. However, if such
small values are approached, inverse-triad corrections become significant and
suppress spatial derivatives. The evolution then follows a nearly homogeneous
behavior of Bianchi-I type, for which singularity resolution in loop quantum
cosmology can be shown in general terms by quantum hyperbolicity
\cite{Sing,IsoCosmo,HomCosmo,BSCG}, based on properties of difference
equations for wave functions. Even without symmetry assumptions and without
restricting the class of quantum corrections included, the dynamics of loop
quantum gravity is singularity-free. The same mechanism is hereby shown to
apply in symmetric models \cite{Sing,HomCosmo,Spin,SphSymmSing} and the full
theory.

The concrete mechanism is reminiscent of the BKL scenario \cite{BKL} in that
spatial derivatives are suppressed and the dynamics becomes almost homogeneous
near singularities. The present scenario, however, is much more general. We
need not rely on details of the evolution because it is terms in the effective
action itself that show the suppression. Moreover, the arguments are easily
seen to be independent on what gauge, or spatial slicing in the classical
setting, is chosen, because they make use of a consistent and anomaly-free
theory exhibiting general covariance (in a deformed sense). Spatial terms are
suppressed even in the $\{H,H\}$-algebra itself. This feature is also
responsible for the covariance of the mechanism: if $\beta$ is very small,
normal deformations of hypersurfaces, governed by $\{H,H\}$ as in (\ref{HH}),
do not generate spatial displacements from $D$. With the suppression by small
$\beta$, normal deformations form a subalgebra of the full
hypersurface-deformation algebra and can be considered in separation,
eliminating the need of homogeneity assumptions.

\subsection{Dispersion relations and causality}

Our results show how consistent deformations of the type (\ref{HH}), for which
several examples have been found in models of loop quantum gravity as recalled
in Section \ref{s:Overview}, affect the form of action principles
reconstructed from them. From this perspective, the universal modification ---
irrespective of the precise form of the correction function $\beta$ --- is
that a new coefficient $\beta$ rescales time derivatives relative to spatial
derivatives in matter terms as well as gravitational ones. The usual
d'Alembertian $\Box=-\partial_t^2+ g^{ij}\partial_i\partial_j$ is replaced by
$\Box_{\beta}:= -\beta^{-1}\partial_t^2+
g^{ij}\partial_i\partial_j$. Dispersion relations and propagation speeds are
then modified in a compatible way for matter and gravity, as shown explicitly
in the special cases considered in \cite{tensor}. (Counterterms in
perturbative realizations of consistency lead to interesting new effects for
non-propagating modes \cite{ScalarGaugeInv,LoopMuk}.) In particular, while
$\beta\not=1$ implies that speeds of massless modes differ from the classical
speed of light, they all propagate at the same speed as light in space-time
according to deformed relativity. All massless excitations propagate with the
velocity $\sqrt{\beta}$ times the classical speed of light for $\beta>0$. If
$\beta<0$, which is possible for holonomy corrections, the d'Alembertian
changes to a Euclidean-signature Laplacian, and all propagation ceases.

\subsection{Signature change}
\label{s:Sig}

Holonomy corrections cannot easily be analyzed in general terms because their
mixing with higher-curvature corrections requires the latter to be derived in
detail, too. In loop quantum gravity, however, the derivation of
higher-curvature terms or their analog in quantum back-reaction remains
incomplete. But there is one general property of holonomy corrections realized
when they are large and near their maximum value. When this is the case, we
must be careful with the $v$-expansions used. One consequence, fortunately,
can be seen very generally.

\subsubsection{The high-density regime in models of loop quantum gravity}
\label{s:HighDens}

In existing consistent examples, holonomy corrections always have the
following form: A connection or extrinsic-curvature component in the classical
Hamiltonian constraint is replaced by a bounded and periodic function of the
same component (possibly depending also on the triad).  For instance, in
spherical symmetry we can consistently replace $K_{\varphi}$ by
$\delta^{-1}\sin(\delta K_{\varphi})$ with some parameter $\delta$ \cite{JR},
and in isotropic models we can replace the isotropic connection component $c$
by $\delta^{-1}\sin(\delta c)$ \cite{ScalarHol}. The parameter $\delta$ may
depend on triad components $E^x$ or $a$ if lattice-refinement is realized
\cite{InhomLattice,CosConst}.  When these bounded functions take their maximum
value, at $\delta K_{\varphi}=\pi/2$ or $\delta c=\pi/2$, holonomy corrections
are large and the Hamiltonian constraint ensures that we are at high energy
densities if matter is present. As recalled in Sec.~\ref{s:Overview}, in the
constraint algebra we obtain a deformation with correction function
$\beta(K_{\varphi})= \cos(2\delta K_{\varphi})$ and $\beta(c)=\cos(2\delta
c)$, respectively.

These functions are negative when $\sin(\delta c)/\delta$ is near its
maximum as a function of $c$, continuing with the example of nearly
isotropic cosmology. More precisely, a modified Hamiltonian constraint
of the form
\begin{equation} \label{modConstr}
 -\frac{3}{8\pi G\gamma^2\delta^2} \sin^2(\delta
c)\sqrt{|p|}+H_{\rm matter}=H\,,
\end{equation}
as commonly used in isotropic loop quantum cosmology, implies, using
$\{c,p\}=8\pi\gamma G/3$, Hamiltonian equations $\dot{p}=\{p,H\}=
(\gamma\delta)^{-1} \sin(2\delta c) \sqrt{|p|}$ and
\[
 \dot{c}= -\frac{\sin^2(\delta c)}{2\gamma\delta^2\sqrt{|p|}}
-\frac{c\sin(2\delta c)}{\gamma\delta \sqrt{|p}} \frac{{\rm d}\log\delta}{{\rm
    d}\log p} + \frac{2\sin^2\delta c}{\gamma\delta^2\sqrt{|p|}} \frac{{\rm
    d}\log\delta}{{\rm d}\log p}
+\frac{8}{3}\pi\gamma G \frac{\partial H_{\rm matter}}{\partial p}\,.
\]
(With $\partial H_{\rm matter}/\partial p= \frac{3}{2}a \partial H_{\rm
  matter}/\partial a^3= -\frac{3}{2}aP$, the usual pressure contribution
$-4\pi G P$ to acceleration follows.) We can combine these two equations to
compute the acceleration of the scale factor, 
\begin{equation}
 \ddot{a}= -\cos(2\delta c) \frac{\sin^2\delta c}{2\gamma^2\delta^2\sqrt{|p|}}
 -\frac{2\sin^4\delta c}{\gamma^2\delta^2\sqrt{|p|}}
  \frac{{\rm d}\log\delta}{{\rm d}\log p} 
-4\pi G\cos(2\delta c)   a P\,.
\end{equation}
To distinguish different types of inflation, it
is also useful to rewrite the acceleration equation as an equation for the
derivative of the Hubble parameter ${\cal H}$:
\begin{equation}
 \dot{\cal H}= \frac{\ddot{a}}{a}-\left(\frac{\dot{a}}{a}\right)^2=
 -\cos(2\delta c)\frac{3\sin^2\delta c}{2\gamma^2\delta^2|p|}
-\frac{\sin^4\delta c}{\gamma^2\delta^2|p|} 
\left(1+2\frac{{\rm d}\log\delta}{{\rm d}\log p}\right)
- 4\pi G\cos(2\delta c) P\,.
\end{equation}

If we assume a power-law form $\delta(p)=|p|^x$ with $-1/2<x<0$
generically \cite{InhomLattice,CosConst}, the gravitational
contributions to $\ddot{a}$ are positive, implying inflation from
quantum geometry, if $\sin^2\delta c>(2(1-2x))^{-1}$ ($\sin^2\delta
c>1/4$ or $\delta c>\pi/6$ for the limiting case $x=-1/2$ considered
in \cite{APSII}). We have super-inflation with $\dot{\cal H}>0$ if
$\sin^2\delta c>3/(4(1-x))$ ($\sin^2\delta c>1/2$ or $\delta c>\pi/4$
for $x=-1/2$). In terms of densities, according to the modified
constraint equation (\ref{modConstr}) showing that the energy density
$\rho$ is proportional to $\sin^2(\delta c)$, we have the maximum
density $\rho_{\rm max}$ when $\sin^2\delta c=1$, inflation for
$\rho>\rho_{\rm max}/(2(1-2x))$ and super-inflation for
$\rho>3\rho_{\rm max}/(4(1-x))$. (For $x\not=-1/2$, $\rho_{\rm max}$
depends on the dynamical discreteness scale $a\delta$.) During
super-inflation, we always have $\cos(2\delta c)<0$, and for $x=-1/2$,
the super-inflationary regime $\sin^2(\delta c)>1/2$ is exactly the
one with $\cos(2\delta c)=1-2\sin^2(\delta c)<0$.

Classically, there can be acceleration only with negative pressure of
a suitable size. But with holonomy corrections, the trigonometrical
factors can turn the sign of $\dot{\cal H}$, providing
matter-independent acceleration from quantum geometry. The correction
function $\beta$ contains the same factor of $\cos(2\delta c)$ that
appears in the acceleration equation. We have a negative correction
function throughout the regime where holonomy effects make $\dot{\cal
H}$ positive, which is in the purported super-inflationary
regime. When $\dot{\cal H}$ is turned positive by holonomy effects, we
therefore do not have space-time but rather (deformed) Euclidean
space, with the derivatives of $a$ taken by spacelike rather than
timelike coordinates. There is no evolution in Euclidean space, and no
super-inflation even if derivatives of ${\cal H}$ are positive. (For
$x<0$, there is still a weak form of power-law inflation at the
beginning of the Lorentzian expansion phase. However, the phase is too
brief, with only a small number of $e$-foldings, for the usual
consequences of inflation to be realized.)

It is of interest to see what an effective action for this Euclidean
chunk of space may look like. In our derivation of effective actions,
applied to such a regime of large curvature, we can no longer expand
the correction function $\beta$ in $K_{\varphi}$ or $c$ when $\delta
K_{\varphi}$ or $\delta c$ is near $\pi/2$, but we can expand them in
$2\delta\bar{K}_{\varphi}:=2\delta K_{\varphi}-\pi$ or
$2\delta\bar{c}:=2\delta c-\pi$, writing $\cos(2\delta K_{\varphi})=
\cos(2\delta \bar{K}_{\varphi}+\pi)=-\cos(2\delta\bar{K}_{\varphi})$
or $\cos(2\delta c)=
\cos(2\delta\bar{c}+\pi)=-\cos(2\delta\bar{c})$. The new coefficients
$\delta \bar{K}_{\varphi}$ or $\delta\bar{c}$ are small near maximum
density, and we can expand the correction function as well as the
Lagrangian by their powers. (For cosmological perturbations around
spatially flat isotropic models, we would expand in $\bar{v}_{ij}:=
v_{ij}-\frac{1}{2}\delta^{-1}\pi\delta_{ij}$.)  Resumming
higher-curvature terms by making use of the small barred quantities,
we obtain effective actions as before. The main consequence of
holonomy corrections then appears even at leading order in the
expansion, for $\ao$ in the new expansion takes the value $\ao=-1$. At
the point of maximum density, where $\delta\bar{K}_{\varphi}=0$ or
$\delta\bar{c}=0$ and therefore $\beta=\ao=-1$, the gravitational
action becomes classical, albeit of Euclidean signature.

\subsubsection{Euclidean space instead of holonomy-induced super-inflation}

Negative $\ao$, in all models studied consistently so far, are a necessary
consequence of holonomy modifications in the high-density phases in which they
may resolve singularities.  With negative $\ao$, however, the dispersion
relation is positive definite and the hypersurface-deformation algebra is of
Euclidean signature, as seen in Sec.~\ref{s:Trans}. These consequences are
consistent with a formal transformation from positive to negative $\beta$ in
(\ref{HH}) by the replacements of $N$ or $t$ by $iN$ or $it$,
respectively. With a Euclidean action, the initial/boundary-value problem
changes its form significantly and propagation in time no longer exists. Loop
quantum gravity, in this way, provides a concrete mechanism for signature
change.

In loop quantum cosmology, going through the Planck regime near the
big bang does therefore not at all correspond to a bounce, as
minisuperspace models are sometimes interpreted as suggesting
\cite{QuantumBigBang}. The big bang is rather a transition from
Euclidean 4-dimensional space to Lorentzian space-time which only
appears dynamical in the homogeneous background. This observation
shows some of the pitfalls and unexpected subtleties of minisuperspace
models. We are also reminded that we have to be careful with
gauge-fixings or deparameterization, which do not determine the
constraint algebra and cannot show the consequences seen here (see
e.g.\ \cite{ScalarHolEv}). One example for difficulties with
deparameterization of cosmological evolution is realized in models
with a positive cosmological constant \cite{HigherMoments}. The range
of internal time provided by a free, massless scalar $\phi$ does not
match with the range of proper time $\tau$ of observers, with $\tau$
diverging at large volume while $\phi$ changes in a finite
range. Extending the internal-time evolution to all values of $\phi$
is then unphysical because no observer could see the extended
space-time solution. The Euclidean phase found here provides another
example, requiring us to bound the range of internal time $\phi$ also
at small volume in loop quantum cosmology. Classically, we know the
space-time structure and all we need to ensure for a good internal
time is that its rate of change ${\rm d}\phi/{\rm d}\tau$ does not
become zero. With a deformed notion of space-time structure, the
derivatives in background equations of motion may not refer to time at
all, and therefore $\phi$ cannot be called an internal ``time'' even
if it keeps changing with the background coordinates. We can start our
internal time $\phi$ only when space-time turns
Lorentzian.\footnote{There can be several reasons for internal times
to be defined for finite ranges only, making them local in
nature. Quantization techniques for totally constrained systems often
require global internal times, but with effective constraints
\cite{EffCons,EffConsRel} there is a systematic framework for quantum
evolution by local internal times
\cite{EffTime,EffTimeLong,EffTimeCosmo}.}

In addition to these cautionary remarks for some scenarios in loop quantum
cosmology, the new picture of signature change also provides larger unity
among the different scenarios for singularity resolution. The main mechanism
\cite{Sing} is based on properties of the underlying difference equations that
appear with a loop quantization \cite{cosmoIV}, with difference operators on
minisuperspace. The resulting recurrence scheme of the wave function depending
on an integer geometrical quantity, taking both signs thanks to orientation,
allows one to evolve uniquely from one side of the classical singularity in
minisuperspace to the other. With unique evolution, the singularity is
resolved in this picture of quantum hyperbolicity making use of geometrical
internal time. A scenario of less generality is realized for deparameterizable
models sourced by a scalar field when its energy is almost all kinetic. Here,
using the scalar as internal time, the minisuperspace evolution is
non-singular with a minimum volume achieved at high density.

These pictures look inconsistent at first sight, with the oriented
volume used as unbounded recurrence variable in the first one, but
bouncing back from a small value in the second one. With the results
of this paper we see that what is inconsistent is not the role of
volume in the recurrence, but rather the interpretation of evolution
as a smooth bounce. In both cases, a collapsing branch of shrinking
volume is connected to an expanding branch of growing volume by a
non-classical space-time region. In the first picture, based on a
recurrence analysis of discrete wave equations, the non-classical
part is modeled as a tunneling process of the wave function through
small volume, while it becomes a Euclidean chunk of 4-dimensional
space in the second picture.  This scenario not only unifies different
mechanisms of singularity resolution in loop quantum cosmology, it
also shows an interesting and unexpected overlap with the tunneling
aspects of \cite{tunneling} and the postulated signature change of
\cite{nobound}.

\subsubsection{The question of cosmological initial values}

We arrive at several new possibilities for cosmological model
building: Initial values can be posed only in the Lorentzian
regime. Holonomy-induced super-inflation, as it appears in the
background evolution in loop quantum cosmology at high density, is not
realized; the corresponding background piece is not part of space-time
but rather corresponds to a Euclidean chunk of 4-dimensional space.
(Super-inflation from inverse-triad corrections
\cite{Inflation,InflationWMAP} has a positive $\beta$ and could happen
in the space-time part.) While the background equations, taken on
their own, might be interpreted as implying super-inflationary
evolution, they fail to provide any insight into the correct
initial/boundary-value problem. Only an extension at least to
perturbative inhomogeneity, without gauge fixing or deparameterization
so as to have access to the off-shell constraint algebra, can provide
this important input, and it shows the Euclidean nature. With the
corresponding boundary-value instead of initial-value problem, even
the background equations can no longer be interpreted as evolution
equations in time.

The Euclidean nature of high-density regimes with holonomy corrections have
several unanticipated consequences for initial values in cosmology. One cannot
use this phase to evolve or generate structure, or to pose initial conditions
within it, such as at the bounce of maximum density. Models making use of the
super-inflationary phase to supply initial values, even if only for the
background equations as suggested for instance in \cite{InflProb}, are not
consistent with quantum geometry.  It becomes, however, very natural to pose
initial values right at the boundary of Euclidean space, cutting off
super-inflation. This procedure would be similar to the usual choice of
initial values or an intial vacuum state before slow-roll inflation, but
providing stronger justification of the choice.

There are several advantages. First, we can pose well-defined initial
values in a non-singular regime. Classically, if we go back as far as
possible to pose initial conditions close to what can be considered
the beginning, we end up at the big-bang singularity. If there is a
bounce \cite{BounceReview}, we end up far back at large volume in the
preceding collapse branch. In the deformed solutions with holonomy
corrections of loop quantum gravity, we end up at the non-singular
beginning of the Lorentzian branch, a clearly distinguished and
non-singular moment in time. Secondly, methods of Euclidean quantum
gravity may be used to shed light on what initial conditions one
should expect. These initial conditions would not be transferred from
the collapse phase bordering the Euclidean chunk at its other end: In
Euclidean 4-space we must choose boundary conditions for a well-posed
formulation of partial differential equations for inhomogeneity. This
boundary includes the initial-value slice of the expanding branch of
the universe model and the final-value surface of the collapsing
branch.  Field values on these surfaces can be specified independently
and freely for a complete set of Euclidean boundary conditions.  We
could, for instance, evolve the collapsing branch from its initial
data to obtain field values at one piece of the Euclidean
boundary. Boundary conditions will then be completed by choosing
values on the rest, including the initial-data surface of the
expanding branch.  Therefore, the final values of the collapse do not
determine initial values for expansion. There is no deterministic
evolution across the Euclidean high-density phase.\footnote{Some
indications of non-deterministic evolution through bounces of loop
quantum cosmology can be found already at the level of background
evolution, where cosmic forgetfulness implies that not all moments of
a pre-big bang state can be recovered after the big bang
\cite{BeforeBB,Harmonic}.} Rather, the scenario describes a
beginningless beginning, with a concrete physical realization of a
distinguished initial-value surface. Although our scenario has cyclic
features in that it combines collapsing and expanding branches,
connected by Euclidean space not causally but at least as manifolds,
we do not encounter the entropy problem. Entropy, like anything else,
will simply not be transmitted through the Euclidean piece.

\subsection{Additional modifications}

Non-local corrections are possible in our formalism, extended from
\cite{LagrangianRegained}, but have not yet been realized explicitly in
effective actions.  We have identified additional difficulties which may
prevent simple realizations of consistent deformations: gravity and matter
terms in the constraints can no longer satisfy the hypersurface-deformation
algebra independently. Instead, there must be delicate cancellations between
matter and gravity Poisson brackets so as to ensure that the total constraints
satisfy a consistently deformed algebra. 

In addition to non-locality, modifications to the spatial part of the
constraint algebra would prevent the steps followed here from going
through. From the perspective of effective constraints, modifications
to the spatial part may not seem likely because these constraints are
formulated for fields on some manifold, which may not obey the
classical geometry but nevertheless is a collection of points labeled,
for the formulation of physical theories, by coordinates. The choice
of coordinates cannot matter for the physics, and so there must be
relabelling invariance. Such an invariance, in turn, leads very
generally to the spatial part of the constraint algebra just based on
properties of the Lie derivative \cite{LagrangianRegained}.

Also from the point of view of full loop quantum gravity,
modifications to the part of the constraint algebra involving the
diffeomorphism constraint may not be called for. This constraint,
unlike the Hamiltonian constraint, is implemented directly by its
action on subsets in space (points or graphs) without any
regularization or modification required to quantize it
consistently. The final verdict on this question has not arrived,
however, as shown by recent attempts to construct diffeomorphism
constraint operators amenable to a closed operator algebras for the
constraints \cite{DiffeoOp}.

The constraint algebra opens the way to specific results for space-time
geometry in loop quantum gravity, extending some minisuperspace results to
more general situations. A crucial open issue remains: deriving consistent
deformations in more general terms than available now. Our results here do not
provide new cases of consistent deformations, because we must assume
consistency in order to employ our algebra. But the new methods do show how
different terms in a consistently modified Hamiltonian constraint must be
related to one another, as seen in conditions for dispersion relations and in
the relations of $v^n$-terms to spatial metric derivatives. Thus, our methods
help in finding new consistent models. But even for existing ones, the
effective actions obtained provide new insights and several unexpected
cosmological consequences.

\section*{Acknowledgements}

We thank Steve Carlip and Bianca Dittrich for discussions about the constraint
algebra, and Thomas Cailleteau and Jakub Mielczarek for emphasizing the
possibility of negative $\beta$. We are grateful to Juan Reyes for sharing
some of his work on consistent constraint algebras in spherical symmetry. This
work was supported in part by NSF grant PHY0748336.


\end{document}